\pdfoutput=1
\documentclass[aps,prd,amsmath,floats,floatfix, twocolumn,
superscriptaddress,nofootinbib,showpacs]{revtex4-1}

\usepackage{amssymb}
\usepackage{amsmath}
\usepackage{graphicx}
\usepackage{dcolumn}
\usepackage{hyperref}
\usepackage{color,units}
\usepackage{lineno}
\usepackage{xspace}
\usepackage{mathtools}
\usepackage{physics}
\usepackage{tensor}
\usepackage[utf8]{inputenc}
\usepackage{acronym}
\usepackage[normalem]{ulem} 

\newcommand{\msun}{\text{M}_{\odot}\xspace}

\newcommand{\AEI}{\affiliation{Max Planck Institute for Gravitational Physics (Albert Einstein Institute), Am M\"uhlenberg 1, Potsdam 14476, Germany}}
\newcommand{\URIPhysics}{\affiliation{Department of Physics, East Hall, University of Rhode Island, Kingston, RI 02881, USA}}
\newcommand{\URICCR}{\affiliation{URI Center for Computational Research, Tyler Hall, University of Rhode Island, Kingston, RI 02881, USA}}

\newcommand{\UMassDMath}{\affiliation{Department of Mathematics,
    University of Massachusetts, Dartmouth, MA 02747, USA}}
\newcommand{\CSCVR}{\affiliation{Center for Scientific Computing and Data Science Research,
    University of Massachusetts, Dartmouth, MA 02747, USA}}
\newcommand{\UU}{\affiliation{Institute for Gravitational and Subatomic Physics (GRASP), Utrecht University, Princetonplein~1, 3584 CC, Utrecht, The Netherlands}}

\begin{document}

\title{A fully precessing higher-mode surrogate model of effective-one-body waveforms}

\author{Bhooshan Gadre}
\email{b.u.gadre@uu.nl}
\AEI
\UU

\author{Michael P\"urrer}
\email{mpuerrer@uri.edu}
\AEI
\URIPhysics
\URICCR

\author{Scott E. Field}
\UMassDMath
\CSCVR

\author{Serguei Ossokine}
\AEI

\author{Vijay Varma}
\thanks{Marie Curie Fellow}
\UMassDMath
\CSCVR
\AEI

\date{\today}

\begin{abstract}
We present a surrogate model of \texttt{SEOBNRv4PHM}, a fully precessing
time-domain effective-one-body (EOB) waveform model including subdominant modes.
We follow an approach similar to that used to build recent numerical relativity
surrogate models. Our surrogate is 5000M in duration, covers mass-ratios up to
1:20 and dimensionless spin magnitudes up to 0.8. Validating the surrogate
against an independent test set we find that the median mismatch error is less
than $10^{-3}$, which is typically smaller than the modeling error of
\texttt{SEOBNRv4PHM} itself.  At high total mass, a few percent of
configurations can exceed mismatches of $10^{-2}$ if they are highly precessing
and exceed a mass ratio of 1:4. This surrogate is nearly two orders of magnitude
faster than the underlying time-domain \texttt{SEOBNRv4PHM} model and can be
evaluated in $\sim 50$ ms.  Bayesian inference analyses with
\texttt{SEOBNRv4PHM} are typically very computationally demanding and can take
from weeks to months to complete. The two-order of magnitude speedup attained by
our surrogate model enables practical parameter estimation analyses with this
waveform family. This is \emph{crucial} because Bayesian inference allows us to
recover the masses and spins of binary black hole mergers given a model of the
emitted gravitational waveform along with a description of the noise.

\end{abstract}


\maketitle

\acrodef{LSC}[LSC]{LIGO Scientific Collaboration}
\acrodef{aLIGO}[aLIGO]{Advanced Laser Interferometer Gravitational wave Observatory}
\acrodef{aVirgo}[aVirgo]{Advanced Virgo}
\acrodef{LIGO}[LIGO]{Laser Interferometer Gravitational-Wave Observatory}
\acrodef{IFO}[IFO]{interferometer}
\acrodef{LHO}[LHO]{LIGO-Hanford}
\acrodef{LLO}[LLO]{LIGO-Livingston}
\acrodef{O2}[O2]{second observing run}
\acrodef{O1}[O1]{first observing run}
\acrodef{BH}[BH]{black hole}
\acrodef{BBH}[BBH]{binary black hole}
\acrodef{BNS}[BNS]{binary neutron star}
\acrodef{NS}[NS]{neutron star}
\acrodef{BHNS}[BHNS]{black hole--neutron star binaries}
\acrodef{NSBH}[NSBH]{neutron star--black hole binary}
\acrodef{PBH}[PBH]{primordial black hole binaries}
\acrodef{CBC}[CBC]{compact binary coalescence}
\acrodef{GW}[GW]{gravitational wave}
\acrodef{CWB}[cWB]{coherent WaveBurst}
\acrodef{SNR}[SNR]{signal-to-noise ratio}
\acrodef{FAR}[FAR]{false alarm rate}
\acrodef{IFAR}[IFAR]{inverse false alarm rate}
\acrodef{FAP}[FAP]{false alarm probability}
\acrodef{PSD}[PSD]{power spectral density}
\acrodef{GR}[GR]{general relativity}
\acrodef{NR}[NR]{numerical relativity}
\acrodef{PN}[PN]{post-Newtonian}
\acrodef{EOB}[EOB]{effective-one-body}
\acrodef{ROM}[ROM]{reduced-order-model}
\acrodef{IMR}[IMR]{inspiral-merger-ringdown}
\acrodef{PDF}[PDF]{probability density function}
\acrodef{PE}[PE]{parameter estimation}
\acrodef{CL}[CL]{credible level}
\acrodef{EOS}[EOS]{equation of state}
\acrodef{LAL}[LAL]{LIGO Algorithm Library}

\newcommand{\PN}[0]{\ac{PN}\xspace}
\newcommand{\CBC}[0]{\ac{CBC}\xspace}
\newcommand{\BBH}[0]{\ac{BBH}\xspace}
\newcommand{\BNS}[0]{\ac{BNS}\xspace}
\newcommand{\BH}[0]{\ac{BH}\xspace}
\newcommand{\NR}[0]{\ac{NR}\xspace}
\newcommand{\GW}[0]{\ac{GW}\xspace}
\newcommand{\SNR}[0]{\ac{SNR}\xspace}
\newcommand{\aLIGO}[0]{\ac{aLIGO}\xspace}
\newcommand{\LIGO}[0]{\ac{LIGO}\xspace}
\newcommand{\Virgo}[0]{Virgo\xspace}
\newcommand{\PE}[0]{\ac{PE}\xspace}
\newcommand{\IMR}[0]{\ac{IMR}\xspace}
\newcommand{\PDF}[0]{\ac{PDF}\xspace}
\newcommand{\GR}[0]{\ac{GR}\xspace}
\newcommand{\PSD}[0]{\ac{PSD}\xspace}
\newcommand{\EOS}[0]{\ac{EOS}\xspace}
\newcommand{\chie}[0]{\chi_{\rm{eff}}\xspace}
\newcommand{\chip}[0]{\chi_{\rm{p}}\xspace}
\newcommand{\seob}[0]{{\texttt{SEOBNRv4PHM}}\xspace}

\section{Introduction}
\label{sec:intro}

The ground-breaking first detection of the merger of a \ac{BBH} system by \LIGO
~\cite{LIGOScientific:2016aoc} in 2015 has opened a new window to the Universe
and heralded the era of \GW astronomy. Since then, the \LIGO and \Virgo detectors
have observed an abundance of merging \acp{BBH}~\cite{LIGOScientific:2016dsl,
PhysRevX.9.031040, PhysRevX.11.021053,LIGOScientific:2021djp}.

To detect and characterize \BBH coalescences we rely on matched filtering
and Bayesian inference~\cite{PhysRevD.44.3819,Allen:2005fk,veitch2015parameter} of the \ac{GW} data recorded by the
interferometer network. This technique requires accurate and fast models of the
emitted gravitational waveform.
While \ac{NR} simulations~\cite{Boyle:2019kee, Ajith:2012az, Aasi:2014tra, GaTechCatalog, Jani:2016wkt,RITCatalog,Healy:2017psd,healy2020third,healy2022fourth}
provide the most accurate waveforms available, they are computationally
expensive, typically requiring anywhere between weeks
to several months per simulation. In addition, \ac{NR} simulations are limited
in duration, and their coverage of the mass-ratio and spin space for precessing
binaries is sparse~\cite{Boyle:2019kee}.

As detector sensitivities improve and the number of observed \acp{BBH} increases,
we are more likely to observe rare and unexpected events such as \acp{BBH} with
asymmetric masses~\cite{LIGOScientific:2020zkf, LIGOScientific:2020stg}, 
or evidence for precession and merger kicks arising from spins misaligned with the orbital
angular momentum~\cite{LIGOScientific:2021djp, Hannam:2021pit, Hoy:2021dqg, Varma:2022pld}, 
or intermediate mass \acp{BH}~\cite{LIGOScientific:2020iuh}.
The wealth of information provided by precessing \acp{CBC} is invaluable to help us
identify astrophysical formation channels
\cite{Farr:2017gtv, Kimball:2020opk}.
Fast and accurate precessing waveform models with subdominant
modes are crucial for extracting information about spin
alignment~\cite{2017PhRvD..95j4038C,Green:2020ptm,PhysRevResearch.2.043096,
Varma:2021csh}.

Thus, to accurately infer the properties of \acp{CBC} with precessing
spins, and binaries with unequal mass-ratios and/or high total mass
where higher harmonics become important, we need reliable waveform
models which can predict the \acp{GW} from such systems.
The effective-one-body (EOB)
approach is a very successful and popular method for constructing such models.
In particular, two different families of EOB models have emerged in recent years,
SEOBNR~\cite{Buonanno:1998gg,Buonanno:2000ef,Pan:2013rra,Babak:2016tgq,Bohe:2016gbl,Steinhoff:2016rfi,Cotesta:2018fcv,Ossokine:2020kjp} and
TEOBResumS~\cite{Damour:2014sva,Nagar:2018zoe,Nagar:2020pcj,Akcay:2020qrj,Gamba:2021ydi}.
Both EOB models include calibration against \ac{NR} and have been shown to compare
well to \ac{NR} simulations not used in their
construction~\cite{albertini2022waveforms,riemenschneider2021assessment,nagar2017impact}.
These models also behave smoothly
outside of their calibration range which makes them useful for the
computation of template banks which need to span a large range of
masses and spins.
In this paper, we focus on applying and extending computational methodology
to accelerate the evaluation time of waveform models that include precession and higher harmonics. For this purpose, we
will mainly focus our efforts on \seob~\cite{Ossokine:2020kjp}, although the techniques presented
here can be readily adapted for other waveform models.

While the \seob model is faithful to NR~\cite{Ossokine:2020kjp}, it is rather expensive to
evaluate which limits its applicability in data analysis. For example, a single likelihood
evaluation is tens of times slower than for waveforms with comparable physics from the Phenom
family~\cite{Estelles:2021gvs, Pratten:2020ceb}. This means that it is challenging to use \seob for parameter
estimation. In particular, standard samplers can have impractical runtimes
even for relatively common events like GW150914: it would take several months to complete
well-converged a run with the \texttt{dynesty} sampler~\cite{Speagle:2019ivv} running on 64 cores~\cite{Smith:2019ucc}.
While it is possible to reduce the wall time by taking advantage of highly parallelized
sampling methods (see e.g~\cite{Smith:2019ucc}) or designing more efficient models
from first principles (see e.g~\cite{Gamba:2021ydi}), computational resources won't be able to
keep pace with the expected large number of detections for future LIGO-Virgo-KAGRA observing runs using slower
models such as \seob.

Another alternative approach is to use approximate grid-based rapid parameter estimation
methods such as RIFT~\cite{Lange:2018pyp,Wysocki:2019grj} that can achieve high efficiency by
evaluating the likelihood on a sparse grid and interpolating it across parameter space. While
these methods have been very successfully used in production analyses~\cite{LIGOScientific:2020zkf,
LIGOScientific:2020iuh,LIGOScientific:2020ibl,LIGOScientific:2021usb}, they often require careful configuration and are hard to robustly scale to the wide variety of expected signals.

To maximize the utility of accurate-but-slow models, such as the \seob model, we require
waveform-acceleration techniques that do not compromise waveform accuracy. The general solution
to this problem is to construct interpolants or fits for waveform data after a dimensionality
reduction step, where the data comes from an accurate underlying model that is typically found
by solving differential equations. This approach has often been referred to as reduced-order or
surrogate modeling~\cite{Field:2013cfa,Purrer:2014fza}, terms that, while technically different,
are by now used interchangeably within the gravitational-wave modeling community. Surrogate models
are accurate in the region of parameter space over which they were trained as well as extremely fast
to evaluate. Surrogate models have been developed to reproduce the radiation from complicated
sources, including signals with $\sim10^4$ waveform cycles~\cite{Purrer:2014fza,Purrer:2015tud,Varma:2018mmi,cotesta2020frequency},
arbitrarily many harmonic modes~\cite{Field:2013cfa,Blackman:2015pia,blackman2017surrogate,
blackman2017numerical,Varma:2018mmi,Varma:2019csw}, spinning binary systems~\cite{Purrer:2015tud,cotesta2020frequency,
Varma:2018mmi}, precessing binary systems~\cite{blackman2017surrogate,Varma:2019csw,
blackman2017numerical}, neutron star inspirals with tidal effects~\cite{lackey2019surrogate,
Lackey:2016krb,barkett2020gravitational}, large- to extreme- mass ratio systems~\cite{chua2019reduced,rifat2020surrogate}, and other diverse problems~\cite{doctor2017statistical,Varma:2018aht,Galley:2016mvy,Bohe:2016gbl,Taylor:2020bmj}. To date, surrogates have only been built for aligned-spin EOB models~\cite{Purrer:2014fza,Purrer:2015tud,Bohe:2016gbl,lackey2019surrogate,cotesta2020frequency,Schmidt:2020yuu}. In this paper, we build a surrogate for the precessing model \seob. Since there are, at present, no surrogate-modeling
techniques for frequency-domain models of generically precessing systems we focus on time-domain techniques previously used for NR
models~\cite{blackman2017numerical,Varma:2019csw}.

We have organized the paper as follows. In Sec.~\ref{sec:methods}, we summarize the surrogate
construction methodology and define the subdomains on which we construct the precessing surrogate.
In the same section, we also describe the detailed choices we make for this particular surrogate
construction. Results and accuracy of the new surrogate model, including its application for
parameter estimation, are discussed in Sec.~\ref{sec:results}. We devote Sec.~\ref{sec:conclusion}
to summarize our conclusions.

\section{Methodology}
\label{sec:methods}

\subsection{Waveform model} 
\label{sub:waveform_and_conventions}

In this section we briefly review the features of the time-domain  model that is used in
constructing the surrogate, \seob.  This model is built using the \ac{EOB} approach,
where the the two-body dynamics is mapped into the motion of a reduced mass in an effective
metric. Analytical information from several sources, such as PN theory, gravitational
self-force, and BH perturbation theory is included in a resummed form. Results from NR
simulations that accurately model the late-inspiral and highly dynamical merger regime are incorporated
into the EOB framework via a calibration procedure.

\seob is a quasi-circular BBH model that includes precession and
modes beyond the dominant quadrupole. This model is based on the aligned-spin multipolar
model \texttt{SEOBNRv4HM}~\cite{Bohe:2016gbl,Cotesta:2018fcv} and is calibrated to 174 NR simulations
in that regime. The model first computes the dynamics  with full spin degrees of freedom. The dynamics is then used to construct a time-dependent co-precessing frame~\cite{Schmidt:2010it,Boyle:2011gg} in which waveform modes are obtained by using aligned-spin expressions from \texttt{SEOBNRv4HM} with spins projected onto the orbital angular momentum at every point in time. The final precessing waveforms are obtained by rotating the co-precessing frame modes back to the inertial frame.

\seob has the spin-2 weighted spherical harmonics (denoted by ${\ell, m}$)
$(2,\pm 2)$, $(2,\pm 1)$, $(3,\pm 3)$, $(4,\pm 4)$, $(5,\pm 5)$ modes in the co-precessing frame and
enforces the conjugate symmetry $h_{\ell, -m} = (-1)^{\ell}h_{\ell, m}^{*}$ in this frame.
While these are considered to be the most important modes, for certain systems and high
signal-to-noise ratios, neglected modes such as the $m=0$ one
can make a non-negligible contribution to the
polarizations~\cite{Gamba:2021ydi}.
While SEOBNRv4PHM sets the $m=0$ modes to zero in the co-precessing frame,
as is done by all other semi-analytic precessing
approximants~\cite{Nagar:2020pcj,Akcay:2020qrj,Pratten:2020ceb,Estelles:2021gvs},
the surrogate modeling methodology can accurately handle the $m=0$ harmonic modes once they are included
in the underlying waveform model~\cite{yoo2023numerical}.

The waveform model takes as input the masses of the compact objects
and Cartesian components of the spins which are defined in the
\emph{source frame}
$\{\hat{e}_{x},\hat{e}_{y},\hat{e}_{z}\}$~\cite{Schmidt:2017btt}. This
frame is constructed at a particular time $t_{\rm ref}$ (which is the
start of the dynamics evolution) such that $\hat{e}_{z}$ is along the
\emph{Newtonian} orbital angular momentum $\hat{L}_{N}$, $\hat{e}_{x}$
is along the $\hat{n}$ vector pointing from secondary to primary and
$\hat{e}_{y}$ completes the triad. It is important to note that the
same physical configuration would have different spin components in
the source frame defined at two different reference times.

The waveform modes are constructed in several steps: i) the dynamics equations are
integrated numerically to produce time-evolutions of the dynamical variables, ii) the
dynamics is used in the construction of the EOB co-precessing frame, and the co-precessing
frame inspiral modes, iii) the merger and ringdown are attached in the co-precessing frame
iv) the waveform modes are rotated back to the intertial source frame.

\subsection{Surrogate construction method} 
\label{sub:surrogate_construction_method}

Our surrogate model is built using a combination of methodologies proposed in previous
works~\cite{Field:2013cfa,Purrer:2014,Purrer:2014fza,Blackman:2015pia,blackman2017numerical,blackman2017surrogate,Varma:2019csw},
which we briefly summarize here.

The goal of a surrogate model is to take a pre-computed {\em training set}
of waveform modes $\{h^{\ell, m}(t;\vec{\lambda}_i)\}_{i=1}^N$ at a fixed set of $N$
points in parameter space $\{\vec{\lambda}_i\}_{i=1}^N$,
and to produce waveform modes $h^{\ell, m}_{\tt S}(t; \vec{\lambda})$
at new parameter values. Here and throughout,
the ``{\tt S}" subscript denotes a
surrogate model for the true data. This modeling task is made
easier by decomposing each waveform into many
\emph{waveform data pieces}. Our strategy is closely
based on the one proposed in~\cite{blackman2017numerical}.
In particular, we model the waveform modes
in a coorbital frame and parameterize
each mode using the instantaneous value of the spin, $\vec{\chi}^\mathrm{coorb}_i(t)$,
in this frame:
\begin{align} \label{eq:waveform_data}
h^{\ell, m}(t; q, \vec{\chi}_1, \vec{\chi}_2 ) \leftrightarrow
h^{\ell, m}_\mathrm{coorb}(t; q, \vec{\chi}^\mathrm{coorb}_1(t), \vec{\chi}^\mathrm{coorb}_2(t)) \,.
\end{align}
Differing from~\cite{blackman2017numerical}, we directly model $h^{\ell, m}_\mathrm{coorb}$ instead of forming
symmetric and anti-symmetric combinations; in the twist-up approximation used by the \seob waveform model, the asymmetric modes are identically zero.

The two-way arrow in Eq.~\eqref{eq:waveform_data}
denotes a map between two descriptions of the waveform data.
On the left-hand-side, we have the standard way of describing
gravitational waveform data where the inertial-frame modes
are parameterized by
the mass ratio and spins measured at a reference time or frequency.
On the right-hand-side, we have the waveform modes that are directly modeled.
To go between these two representations, we also build
a {\em dynamics surrogate model} that includes the following
additional data pieces: the orbital phase, $\phi(t)$, the
unit quaternion that defines the coprecessing frame~\cite{Schmidt2010,OShaughnessy2011,Boyle:2011gg}, $q(t)$, and the
spins in the coorbital frame, $\vec{\chi}^\mathrm{coorb}_{1,2}(t)$.
This is done using the transformation $T_C$ given by Eq.~25 of Ref.~\cite{blackman2017surrogate}.

The individual spins of the two compact objects are naturally only available in the \seob model before merger. Therefore, to continue to have two spins present after the merger, we follow the approach used in NR surrogates~\cite{Varma:2019csw} and artificially extend the component spins using 2~\PN spin evolution equation as described in \cite{Ossokine:2015vda}. We emphasize that these spins do not have a physical meaning, but are simply used as inputs to simplify the construction of fits.

For each waveform data piece,
we construct a linear basis using singular value decomposition (SVD) with a prescribed tolerance.
We then
construct an empirical time
interpolant~\cite{Barrault:2004,Maday:2009,chaturantabut2010nonlinear,canizares2013gravitational} with the same
number of empirical time nodes as basis functions for that
data piece.
The set of empirical time nodes $\{T_j\}$ are
chosen differently for each waveform data piece.
Similar to classical
interpolation (e.g., polynomial basis functions with Chebyshev nodes), using
the SVD basis and set of empirical node times we are able to represent each
data piece with an empirical interpolant.  This interpolant is specific to the
waveform data piece, which takes advantage of a nearly optimal compact
representation of the parameterized time-series
data~\cite{field2011reduced,Field:2013cfa}.

Finally, for each waveform data piece, we construct a \emph{parametric fit}
for the data at time $T_j$, which
is parameterized not by
the spins at some reference time but rather the spins at time
$T_j$. The function $h^{2, 2}_{\mathrm{coorb},{\tt S}}$ is given by a
linear combination of basis functions.
Similar to
Ref.~\cite{blackman2017numerical}, we choose the basis functions to be
a tensor product of 1D monomials in the spin components and $x(q)$,
where $x$ arises from an affine mapping from the interval
$[q_{\rm min}, q_{\rm max}]$ to the standard interval $[-1,1]$.  We
consider up to cubic functions in $x$ and up to quadratic functions in
the spin components.  Using all possible terms of this form would lead
to overfitting and low-quality fits. We correct for this issue by
using the forward-stepwise greedy fitting method described in Appendix
A of Ref.~\cite{blackman2017surrogate}, which selects at most 150 relevant terms in such a way that avoids overfitting.
Increasing the polynomial order in spin components or the mass ratio leads to a very steep increase in the number of terms, making the fit computationally infeasible.


\subsection{Domain decomposition} 
\label{sub:domain_decomposition}

\begin{figure}
  \includegraphics[width=.45\textwidth]{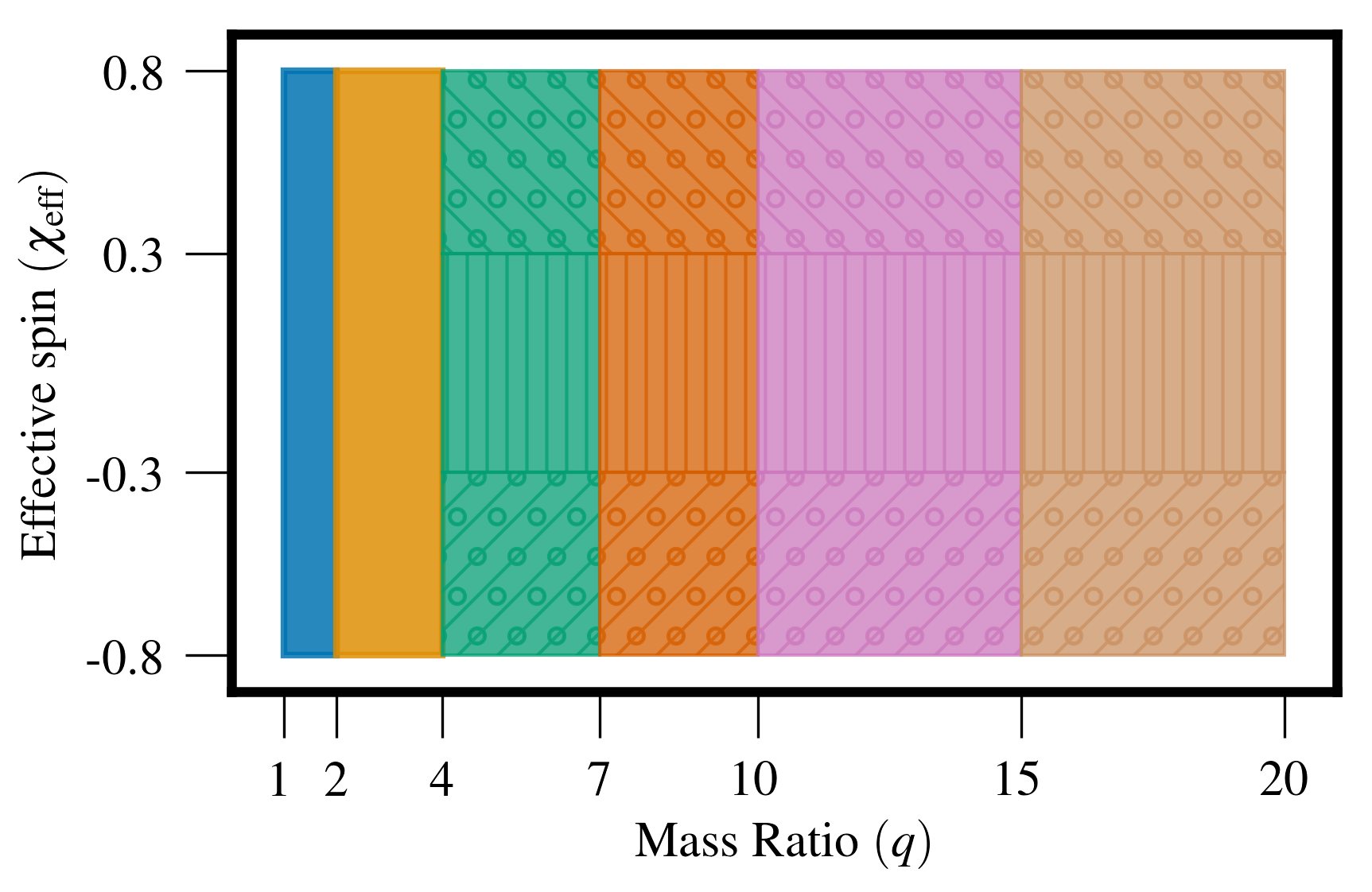}
  \caption{This figure shows the 14 subdomains in mass-ratio $(q)$ and effective spin parameter
  $(\chie)$ of the \seob surrogate. Different colors correspond to the division along $q$ while
  shading types defines boundaries along $\chie$.}
  \label{fig:subdomains}
\end{figure}

Our goal is to build an extensive time-domain surrogate of \seob that can be used to analyze most
of the BBH signals we expect to see with current and near-future ground-based detectors. To
accomplish this, we build our model on a mass ratio interval of $q := m_1 / m_2 \in [1,20]$, which encompasses
all BBH signals detected to date. However, we restrict the spin magnitudes to satisfy $\lvert \chi_{i}
\rvert \leq 0.8$ and place no restrictions on the spin tilts. This follows the choices made
by the recent NR surrogate models~\cite{blackman2017numerical,blackman2017surrogate,Varma:2019csw}
for which high spin simulations are rather sparse. For the \seob model,
where we do not have this restriction,
we have explored building surrogates for spin magnitudes up to $1$. However, we found that surrogate
errors increased rather steeply beyond a spin magnitude of $0.8$ and therefore imposed this spin
bound. We leave an extension of the surrogate up to extremal spins as future work.
We construct the surrogate with a duration of $5000 M$ prior to the merger,
which corresponds to around 1 to 3 precession cycles.

Despite significant experimentation, we were unable to build high-accuracy surrogate models when
attempting a global parametric fit over the full parameter space mentioned above. This is a common
issue for ambitious regression problems, and the general solution is domain decomposition. In the
domain decomposition approach, we break up the large parameter space into smaller subdomains and
build models on each patch. The general expectation is that high-accuracy parametric fits over a
smaller subdomain will be easier to achieve, since variations in the waveform data pieces are
reduced.

We divide our target parameter domain into 14 subdomains in the effective spin parameter, $\chie$,
and mass ratio directions. There are two reasons to choose subdomains in mass ratio and effective
spin. First, the number of cycles in \GW signals of fixed duration depend most strongly on $q$ and
$\chie$. Second, throughout the inspiral $q$ is constant while $\chie$ is nearly
so~\cite{Gerosa:2015tea}. For the mass ratio, we place our domain boundaries at $q = 1, 2, 4, 7,
10, 15,$ and $20$. For $q > 4$, the mass-ratio domains are further divided into 3 subdomains
each in effective spin with boundaries at $\chie = (-0.8, -0.3, 0.3, 0.8)$. The resulting
subdomains are shown in Fig.~\ref{fig:subdomains}. To have a smooth combined surrogate
across the subdomain boundaries, we extend the subdomains so that they overlap by $\pm 0.25$
in $q$ and by $\pm 0.05$ in $\chie$.

\subsection{Choice of time nodes} 
\label{sub:choice_of_time_nodes}
In this section, we discuss the time resolution used for the waveform and dynamics surrogates.
To construct a surrogate for the waveform modes, we choose 2000 time nodes equally spaced in
the orbital phase for a representative waveform in a given subdomain. This choice ensures
an approximately constant number of nodes per orbit from the early inspiral up
through merger (see Appendix B of Ref.~\cite{blackman2017numerical} for details). We build a reduced-order empirical
interpolant representation for each of the waveform modes in the coorbital frame, as described in Sec.~\ref{sub:surrogate_construction_method} with an SVD tolerance of $10^{-5}$.
To build a {\em dynamics surrogate model}, we choose 25 time nodes per orbit. We find that these
particular choices sample the data pieces in time well enough so that we can interpolate this data
to the desired sampling frequency with high accuracy.
%

\subsection{Training and validation set generation}
\label{sub:training_set_generation}

To construct an accurate surrogate, we first need to build a training set that is sufficiently
dense and captures the variations of the data well at each time node. To achieve this, in each subdomain we proceed in three steps:

\begin{enumerate}
  \item We start with 2000 samples in each subdomain uniform in $q$ and for each component spin,
  uniform in a spherical volume with $\lvert \chi_{i} \rvert \le 0.85$. While our target problem is defined on
  $\lvert \chi_{i} \rvert \leq 0.8$, we find that sampling up to $0.85$ will result in models that retain high accuracy near $\lvert \chi_{i} \rvert \approx 0.8$.
  \item Next, we add 2000 sample points chosen uniformly in $q$ and uniformly in $\chie$. To sample
  uniformly in $\chie$, we use rejection sampling on the spin component samples drawn uniformly
  in the spherical volume as above.
  \item Finally, to sample adequately in the effective precession spin parameter $\chip$, we start
  by drawing 8000 points uniform in $q$, $\chip$ and $\chie$, again using rejection sampling, and
  then choose 2000 of the samples which maximize the Euclidean distance in the $\chie - \chip$
  plane.
\end{enumerate}

With the above choice of sampling method, we net a total of 6000 training points per
subdomain, and thus a grand total of 84,000 points for the entire surrogate. The procedure allows
us not only to have an adequate coverage in $\chip$ and $\chie$ (which encode the dominant
precessional and aligned-spin effects), but also provides sufficient samples in the remaining spin
degrees of freedom in the waveform space to capture subdominant spin effects.

In the construction of the training set mentioned above, $\chie$ and $\chip$ are measured at the
reference dimensionless frequency $M\omega = 0.014$. As the constructed surrogate has a duration
of $5000$M until merger, our spin distribution evolves from the one sampled at the reference
frequency to the one measured at the reference time for the surrogate of $-5000$M. The choice of
the surrogate duration roughly corresponds to the waveform's $(2,2)$ mode starting at $20$ Hz when both the component masses are $25 \msun$.

We verify the sufficiency of the chosen training set by validating the surrogate model against an
independent set of validation waveforms sampled using the same procedure as used for the training set.
We reach the final training set by iteratively enriching it with configurations which have
mismatches greater than $1.5$ \% in the validation set. We find that this enrichment improves the
surrogate model only for some of the subdomains.
In particular, the surrogate accuracy can improve if the enriched training set resolves
under-represented features and improves overall goodness of fit. We empirically
found that after two such iterations the model's accuracy no longer improves.

\section{Results}
\label{sec:results}

\subsection{Accuracy} 
\label{sub:accuracy}

In this section we define error measures for surrogate components and study their impact on the
inertial waveform modes generated by the surrogate. We also compute the
noise-weighted Fourier domain matches typically used in gravitational-wave
data analysis.

\subsubsection{Error measures for surrogate data pieces}

The approximation quality of each surrogate data piece is assessed by computing a root-mean-square (RMS)
error
\begin{equation} \label{eq:rms_err}
   \mathcal{E}_{\mathrm{rms}}[X] := \sqrt{\frac{\int_{t_{\min}}^{t_{\max}} |\hat{X}-X_\mathrm{EOB}|^2 ~dt}{T} },
\end{equation}
where $X_\mathrm{EOB}(t)$ is any time dependent waveform data piece taken from the \seob model,
$\hat{X}(t)$ is the surrogate for this data piece, and $T = t_{\max} - t_{\min}$ is the total time duration.

In addition, we consider a relative $L_2$-type error as defined by~\cite{blackman2017surrogate}
(which can be motivated from a white-noise time-domain mismatch)

\begin{equation} \label{eq:abs_err}
    \mathcal{E}_X := \frac{1}{2} \frac{\int_{t_{\min}}^{t_{\max}} \sum_{\ell, m} |\hat{h}_X^{\ell m}-h_\mathrm{EOB}^{\ell m}|^2 ~dt}
                                      {\int_{t_{\min}}^{t_{\max}} \sum_{\ell, m}|h_\mathrm{EOB}^{\ell m}|^2 ~dt}.
\end{equation}
Here $h_\mathrm{EOB}(t)$ is the true inertial frame \seob waveform for a particular BBH configuration.
In contrast, $\hat{h}_X(t)$ is computed using the surrogate waveform decomposition described in
Sec.~\ref{sub:surrogate_construction_method} and the following prescription:
all waveform data pieces are given by the true \seob data except for one particular data piece,
$X$, which is taken from the surrogate for this quantity. This construction allows us to study
how the surrogate modeling errors in a particular data piece (e.g. the orbital phase, the quaternions,
or the coorbital modes) affect the overall inertial modes output by the surrogate.
Furthermore, we also compute $\mathcal{E}_X$ for the
complete surrogate i.e. using only surrogate
data pieces without any true \seob data, as this is the way the surrogate will be used
in practical applications and it contains the total error budget.

\begin{figure*}[htb]
  \includegraphics[width=.49\textwidth]{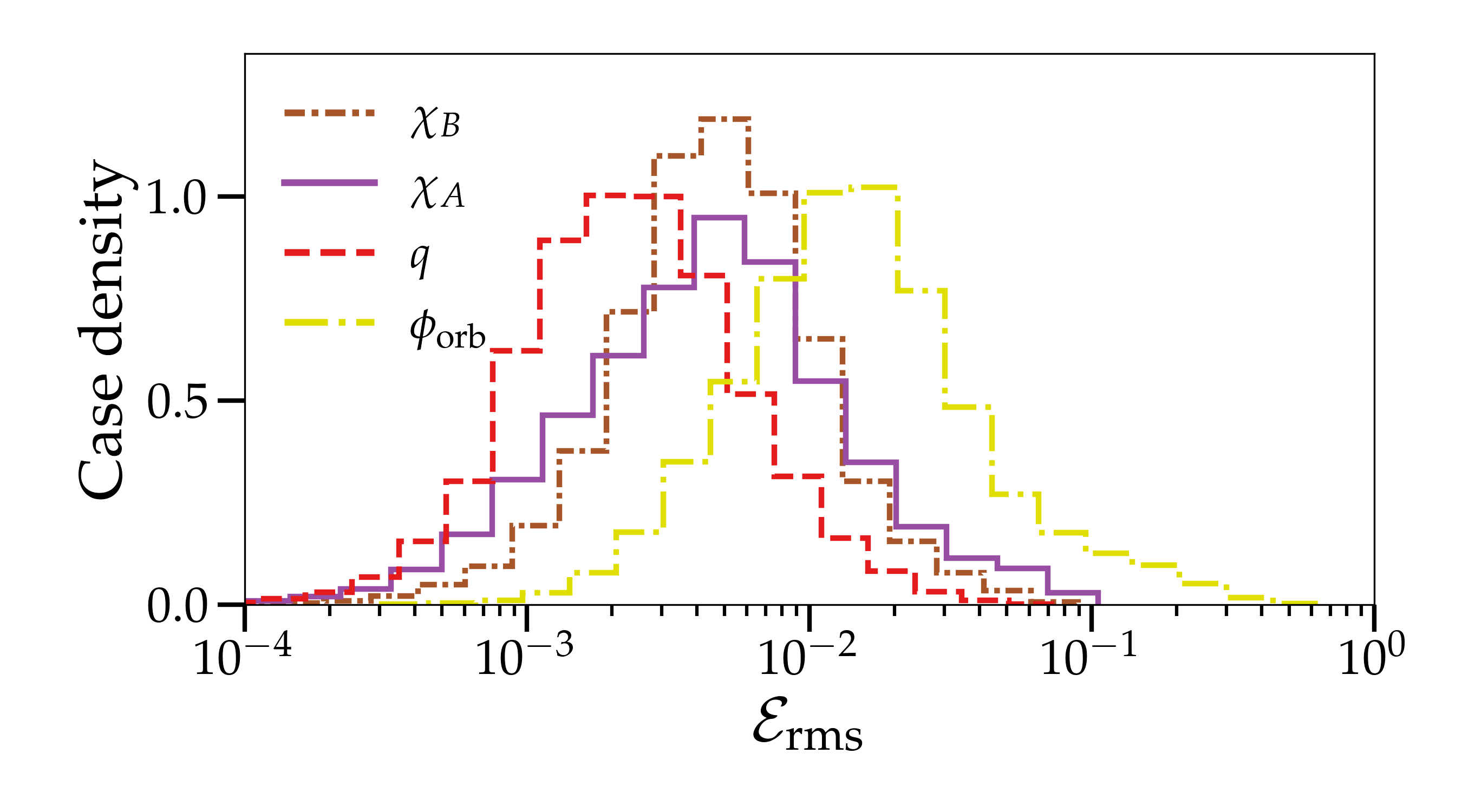}
  \includegraphics[width=.49\textwidth]{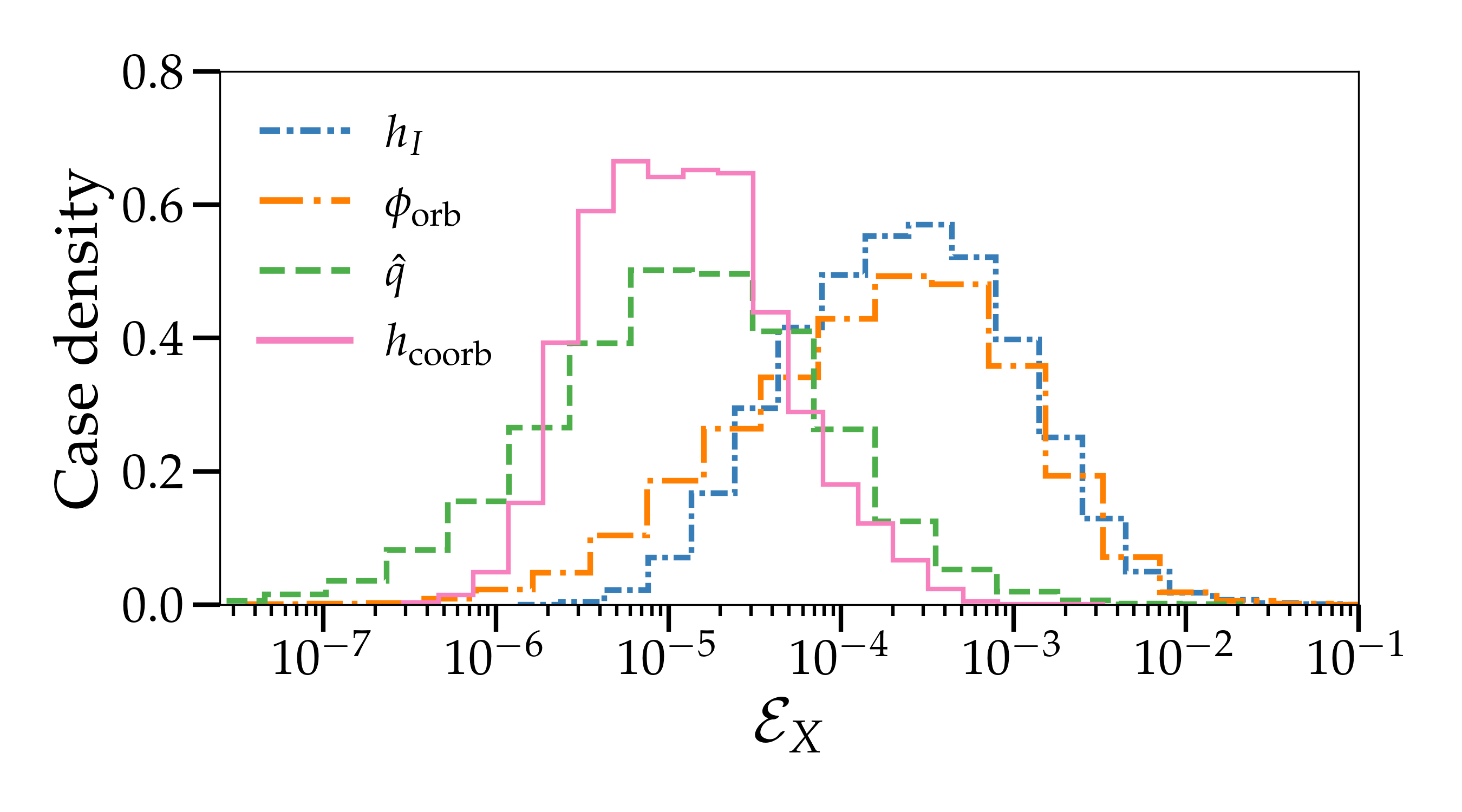}
  \caption{Histograms of errors due to the errors in respective surrogate data pieces on the training set.
    \emph{Left:} RMS errors due to dynamics quantities: spin vectors (solid violet and dash-dotted
    brown), unit quaternions (red dashed), and orbital phase (yellow long dash-dotted);
    \emph{Right:} $L^2$-type errors for the inertial modes due to immediate dependent quantities:
    inertial modes (blue dash-dotted; as as the full surrogate), orbital phase (orange long dash-dotted), quaternions (green
    dashed), and coorbital modes (pink).
  }
  \label{fig:surrogate_train_errors}
\end{figure*}

\begin{figure*}[htb]
  \includegraphics[width=.49\textwidth]{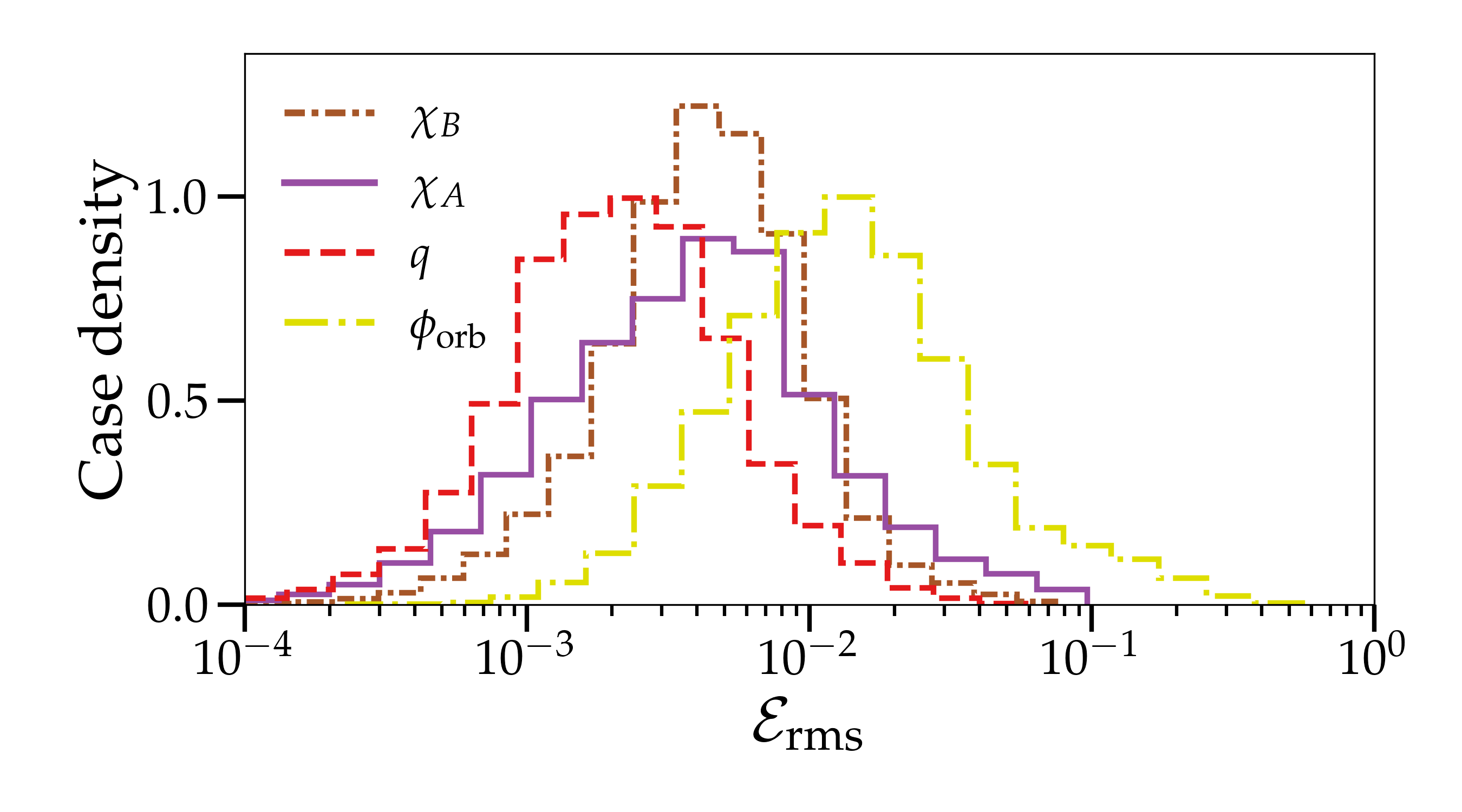}
  \includegraphics[width=.49\textwidth]{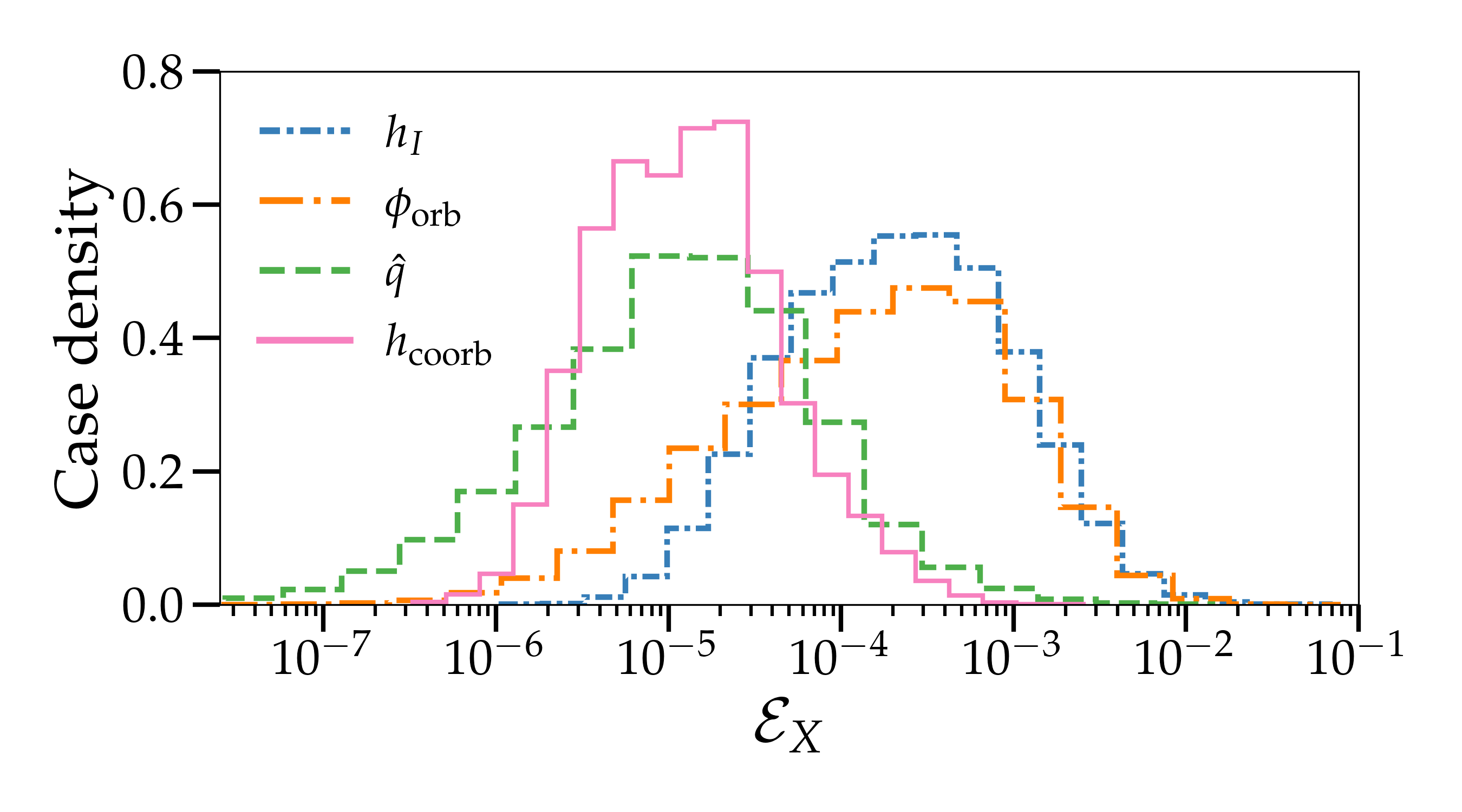}
  \caption{Histograms of errors due to respective surrogate data pieces on the validation set.
    Description of panels and plotted quantities as in Fig.~\ref{fig:surrogate_train_errors}.
  }
  \label{fig:surrogate_test_errors}
\end{figure*}

Fig.~\ref{fig:surrogate_train_errors} shows error histograms for individual data pieces used to
construct the surrogate on the training set and the overall surrogate error.
We can see from both panels that the orbital phase (and hence the orbital frequency) is the source
of the dominant error as compared to the inertial frame waveform errors in the right panel.
This shows that the phase (or frequency) model largely limits the accuracy of the surrogate.
On the validation set which is comprised by more than $84000$ independent test cases
we can see similar behavior as on the training set, as shown in Fig.~\ref{fig:surrogate_test_errors}.

In addition, we can also observe from the $\mathcal{E}_X$-histograms that on both data sets
the coorbital modes and quaternion fits do not limit the surrogate accuracy. The RMS error
histograms suggest that the tails showing large errors in the spin fits do not affect the
coorbital mode and quaternion fits.
To complement the histograms, we note that the surrogate fits worsen as we approach the merger.
This affects the accuracy of the surrogate model for \BBH systems with high total masses.

\subsubsection{Match computations}

We further check the accuracy of the surrogate model by computing matches against the original
\seob waveform model. We use the ``min-max match'' and ``max-max match''
(both denoted by $\mathcal{M}$) as defined by Eq.
B10 in Appendix B of~\cite{Babak:2016tgq} along with the advanced LIGO design
\ac{PSD}~\cite{aLIGODesignNoiseCurve}.
We perform the match computation over the grid of the total mass of the binary and the inclination angles. We use a frequency range between $f_{\rm min}= 25$ Hz and $f_{\rm max}=2048$ Hz. We specify the
initial configuration for \seob waveform generation in the $\vec{L}_{N}$-frame.

To generate the corresponding surrogate waveform and data pieces, we
transform the spin vectors from the $\vec{L}_{N}$-frame to the coprecessing frame which
is the one used for the surrogate model.

\begin{figure}[htbp]
  \includegraphics[width=.48\textwidth]{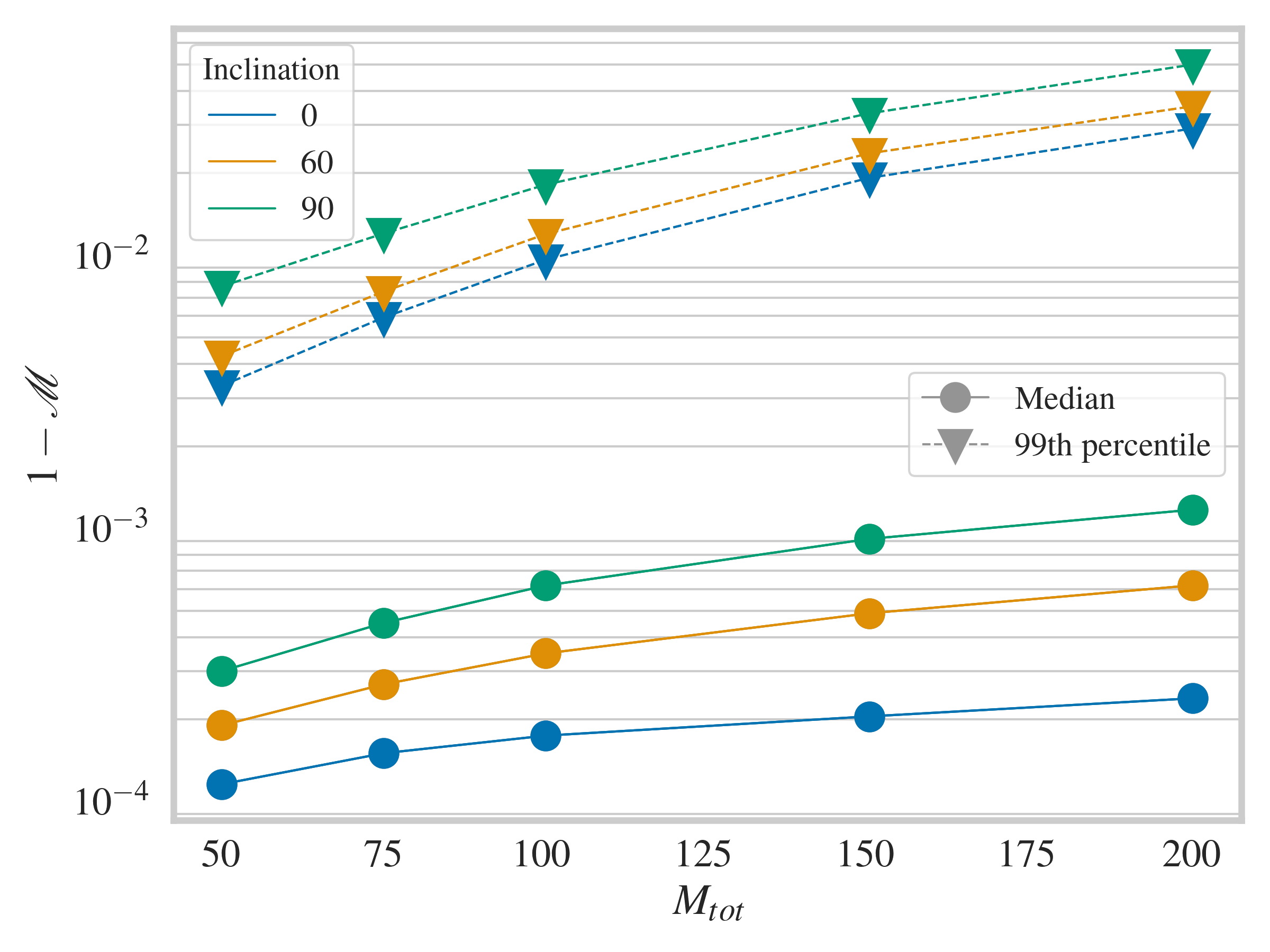}
  \caption{This figure shows mismatches (against min-max match) computed with Advance LIGO design PSD as a function of total mass for 3 different inclinations. Circles and triangles show median and 99th percentile values for training and validation sets combined.}
  \label{fig:all_mismatch_mtot_incl_median}
\end{figure}

\begin{figure}[htbp]
  \includegraphics[width=.48\textwidth]{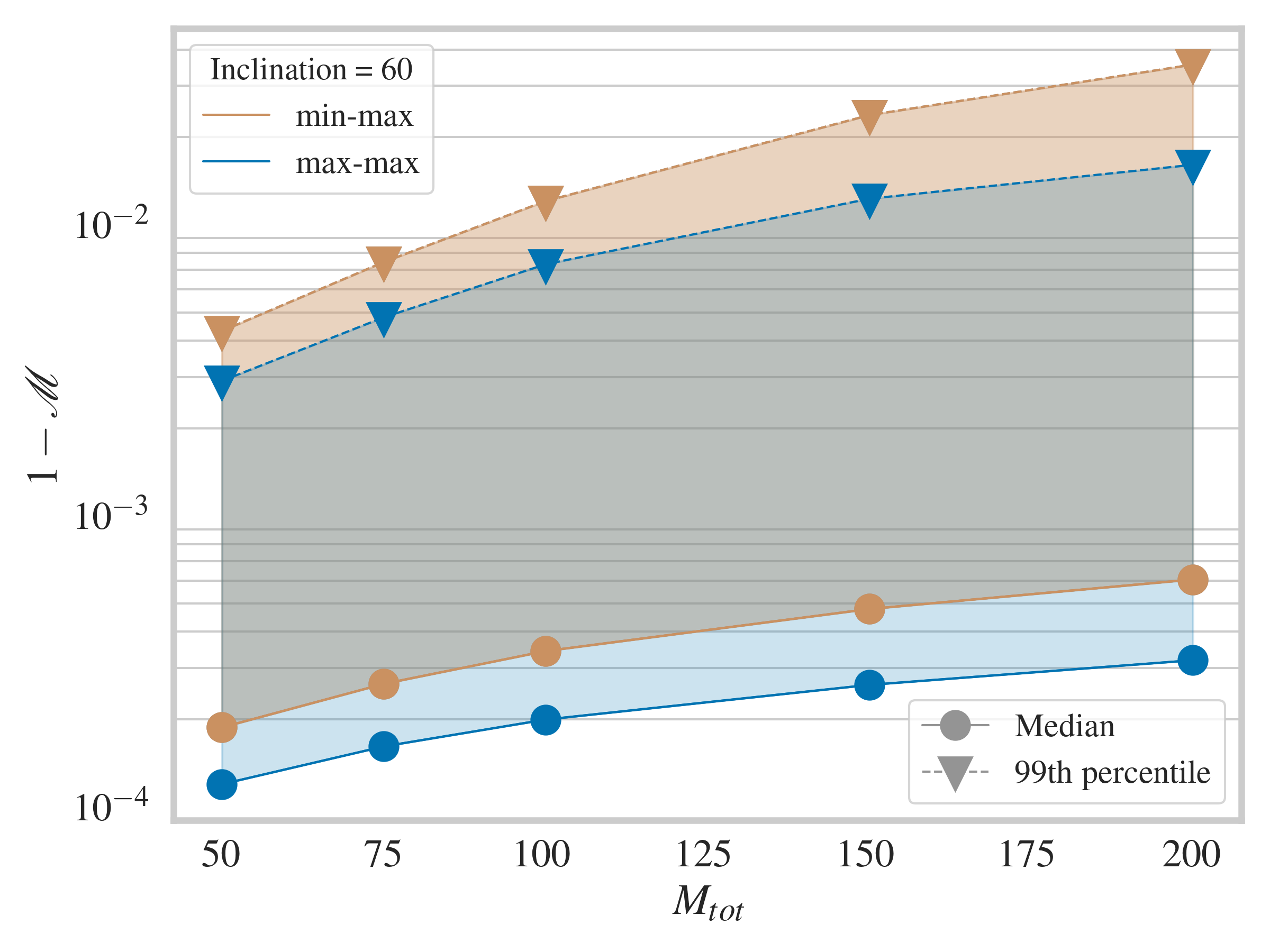}
  \caption{The figure shows representative distribution of mismatches against "min-max" and "max-max" matches computed with Advance LIGO design PSD as a function of total mass for the fixed inclination. The "min-max" and "max-max" matches provide measure of conservative and best case match scenarios respectively.}
  \label{fig:all_mismatch_incl60}
\end{figure}

\begin{figure}[htbp]
  \includegraphics[width=.48\textwidth]{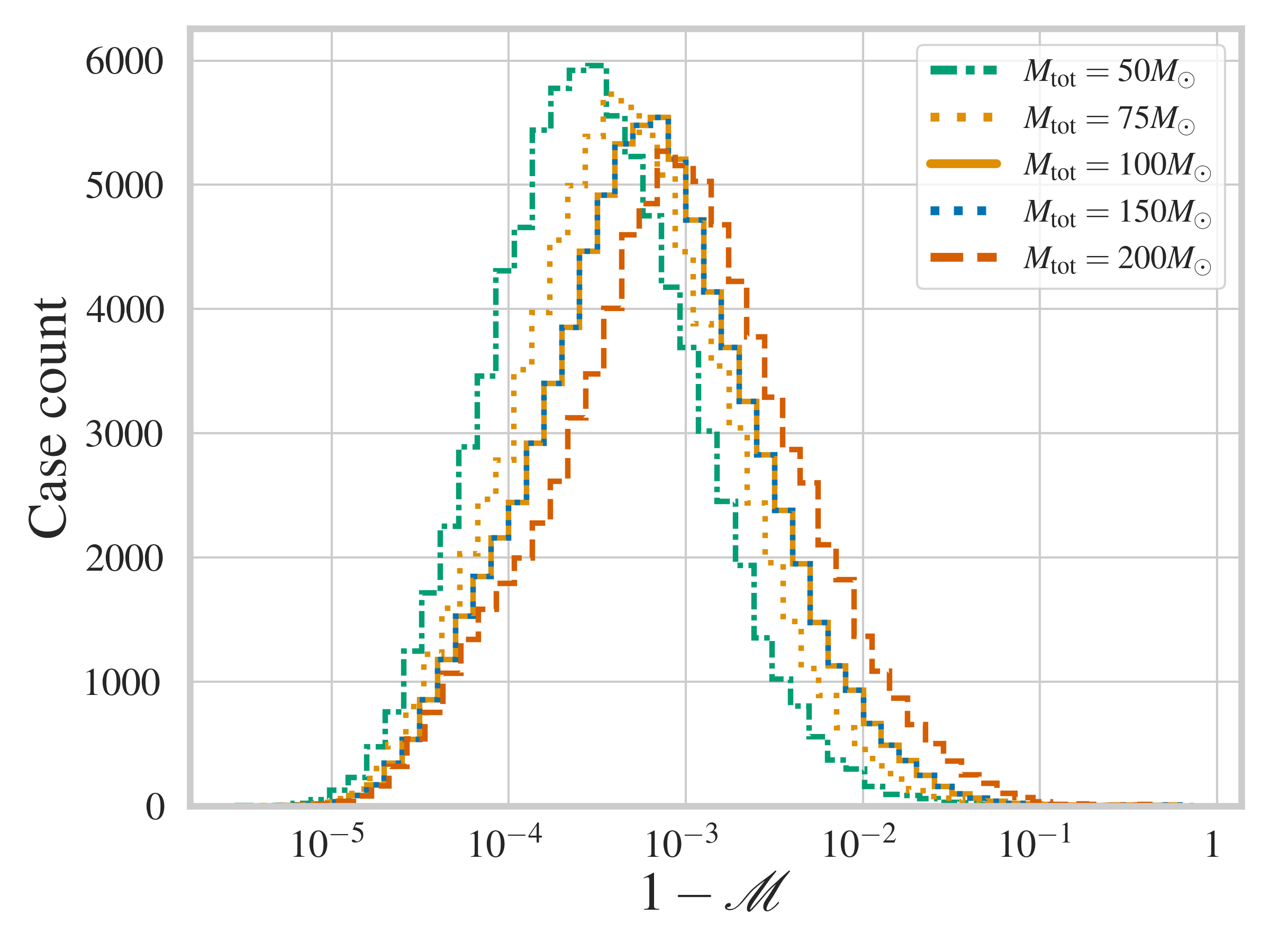}
  \caption{The plot shows the histograms of mismatches between the surrogate
  and the \seob waveform model for validation set for all inclinations for varying total mass.}
  \label{fig:all_match_histogram}
\end{figure}

\begin{figure*}[htb]
    \includegraphics[width=.49\textwidth]{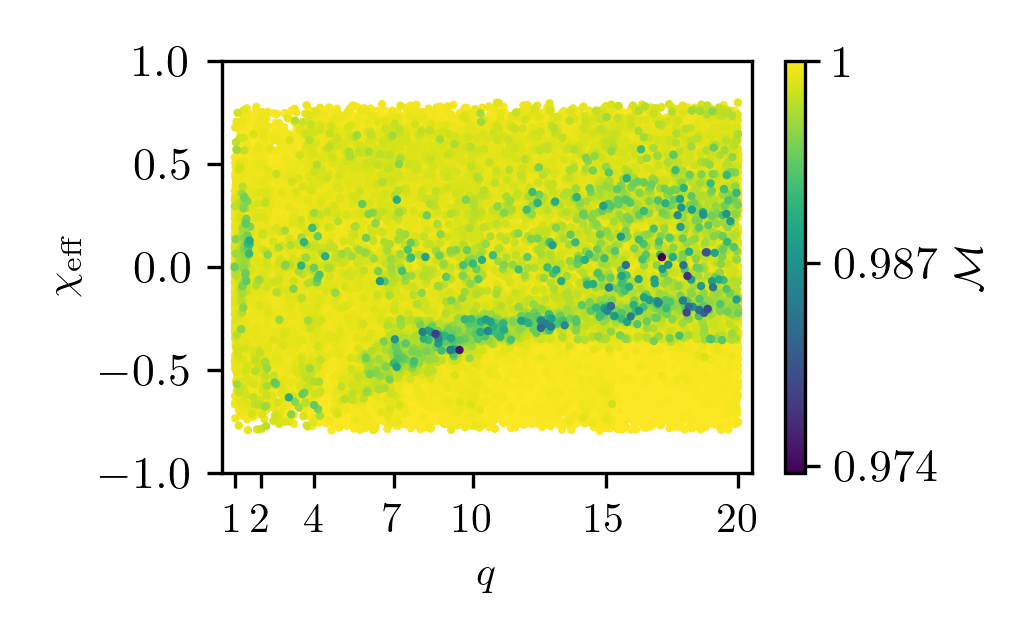}
    \includegraphics[width=.49\textwidth]{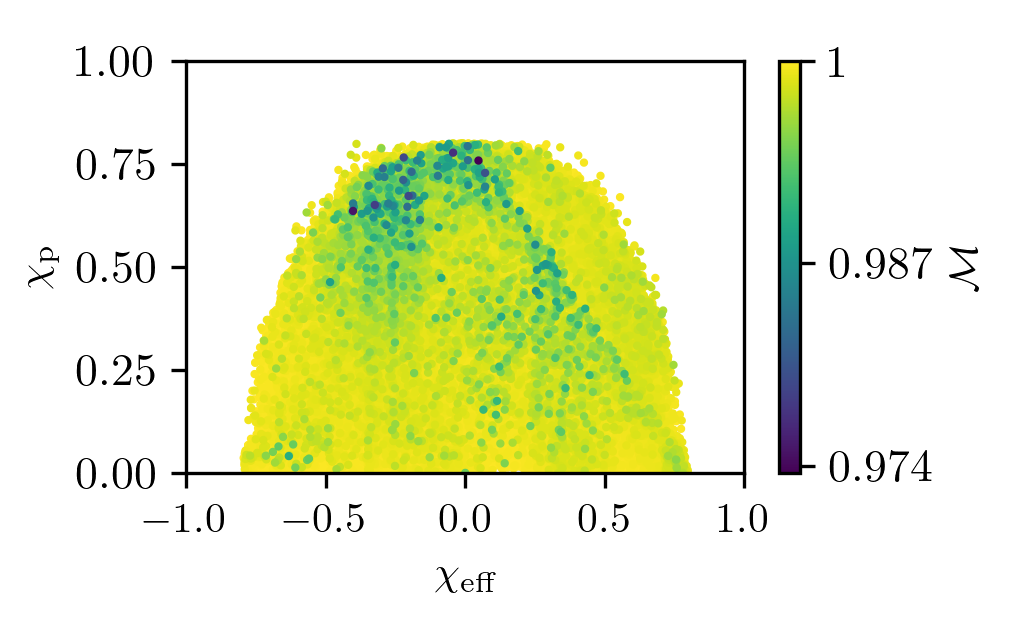}

   \caption{ Scatter plots of matches between the surrogate and underlying precessing EOB waveform model
     for the training set and a total mass of 50 $\msun$.
     \emph{Left:} Matches as a function of mass ratio and effective spin parameter ($\chie$)
     \emph{Right:} Matches in the $\chie$ - $\chip$ plane.
     In both panels the value of the match is encoded by the color of the partially transparent
     circles which are ordered so that the lowest matches (shown in dark green) lie on top of
     higher matches (light yellow).
  }
  \label{fig:train_match}
\end{figure*}

\begin{figure*}[htb]

   \includegraphics[width=.49\textwidth]{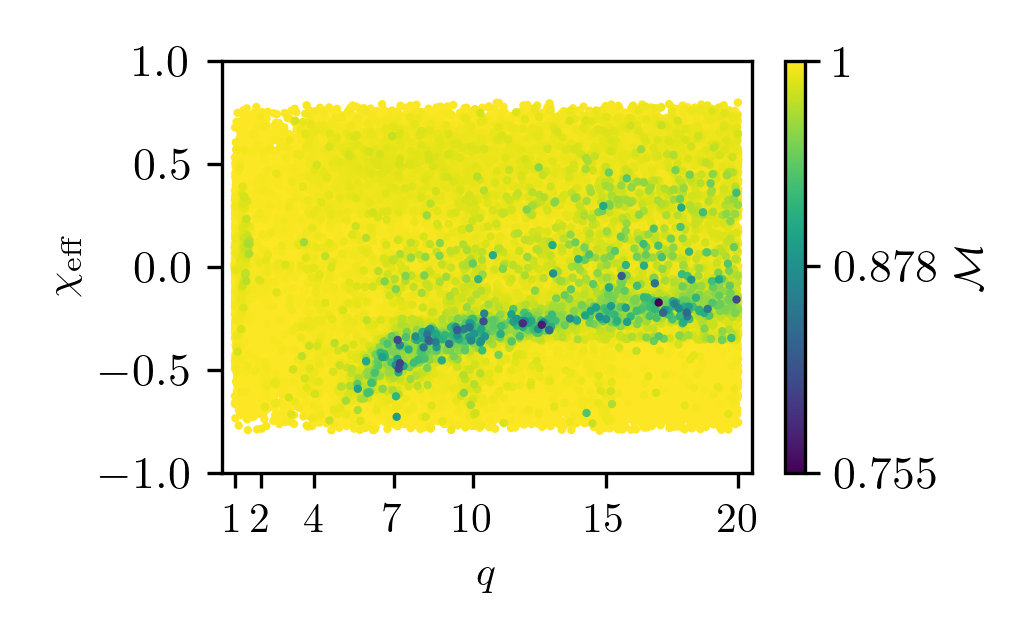}
   \includegraphics[width=.49\textwidth]{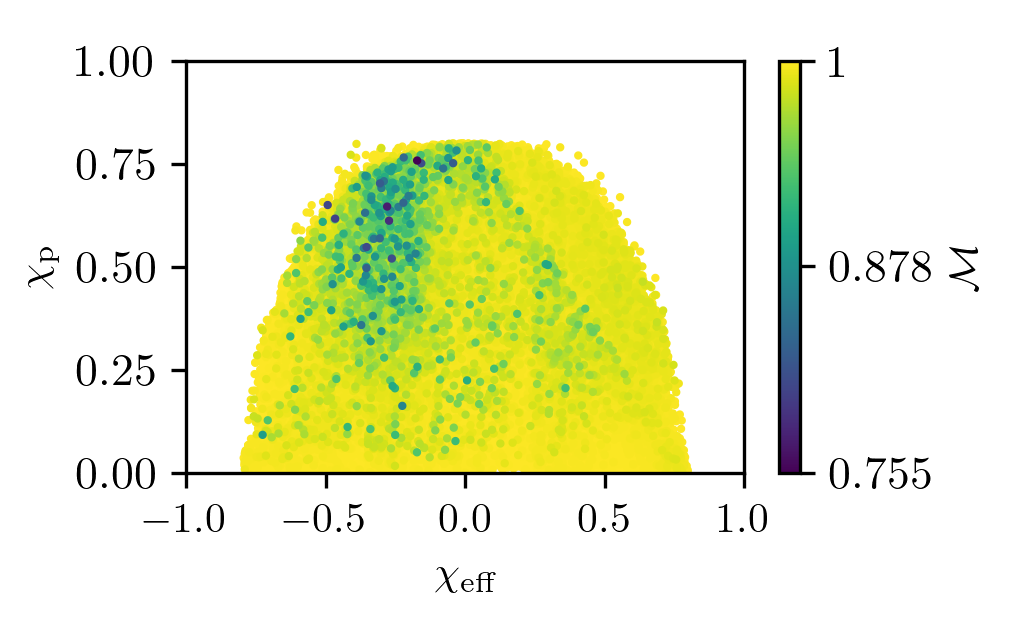}
   \caption{ Scatter plots of matches between the surrogate and underlying precessing EOB waveform model
     for the training set and a total mass of 200 $\msun$.
     The setup of the panels and the shown matches follows Fig.~\ref{fig:train_match}.
  }
  \label{fig:train_match_200}
\end{figure*}

\begin{figure*}[htb]
   \includegraphics[width=.49\textwidth]{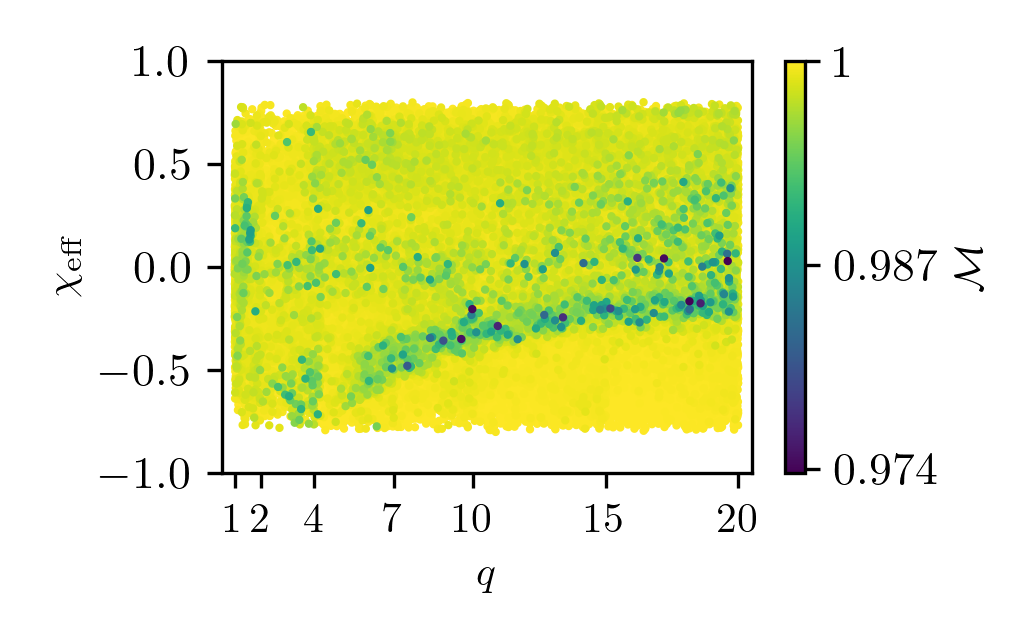}
   \includegraphics[width=.49\textwidth]{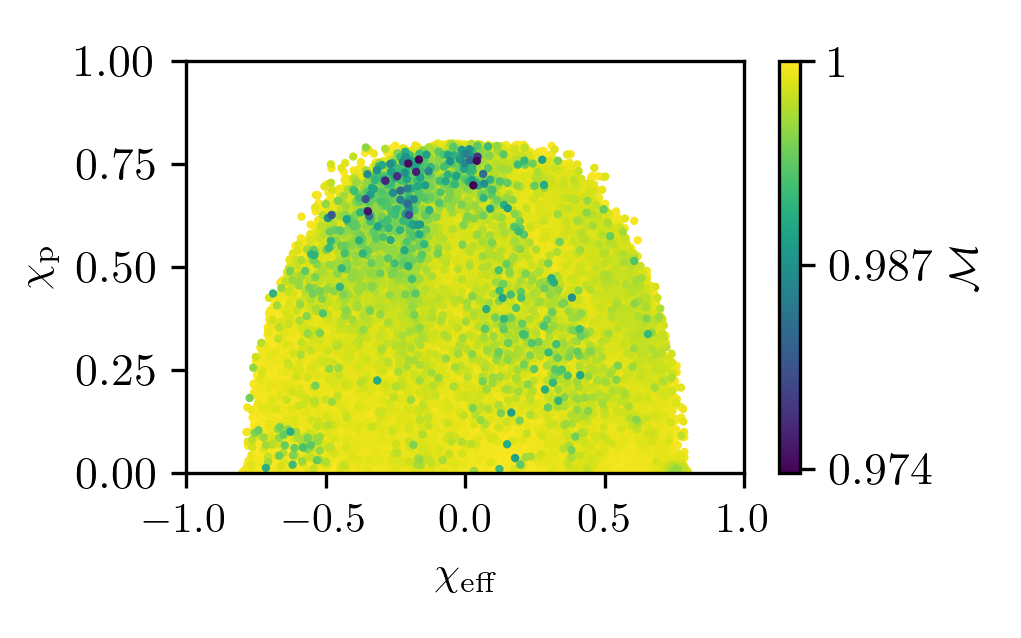}
   \caption{Scatter plots of matches between the surrogate and underlying precessing EOB waveform model
     for the validation set and a total mass of 50 $\msun$.
     The setup of the panels and the shown matches follows Fig.~\ref{fig:train_match}.
  }
  \label{fig:valid_match}
\end{figure*}

\begin{figure*}[htb]

    \includegraphics[width=.49\textwidth]{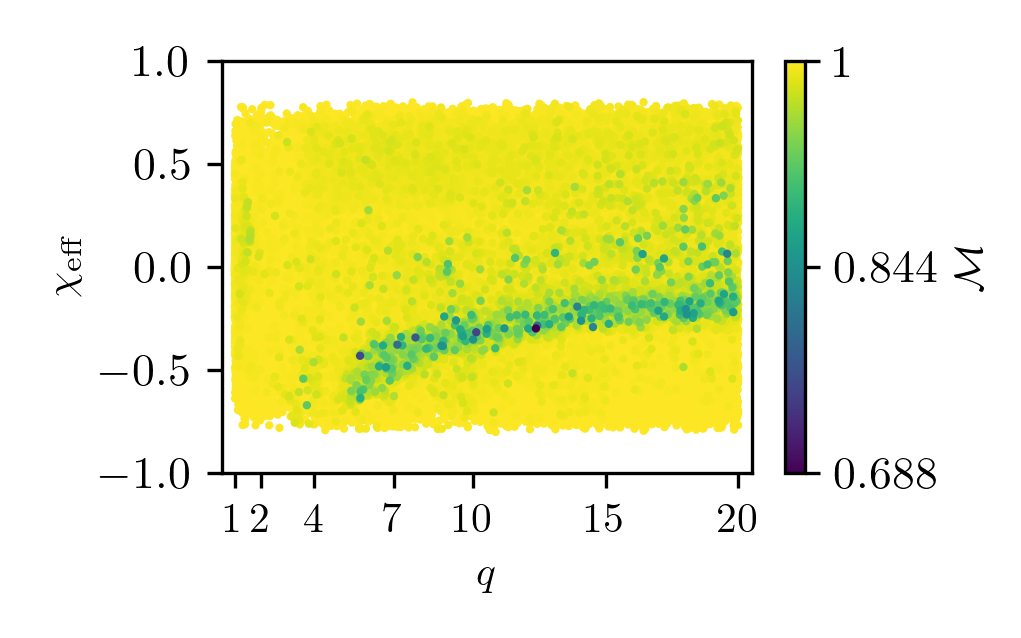}
    \includegraphics[width=.49\textwidth]{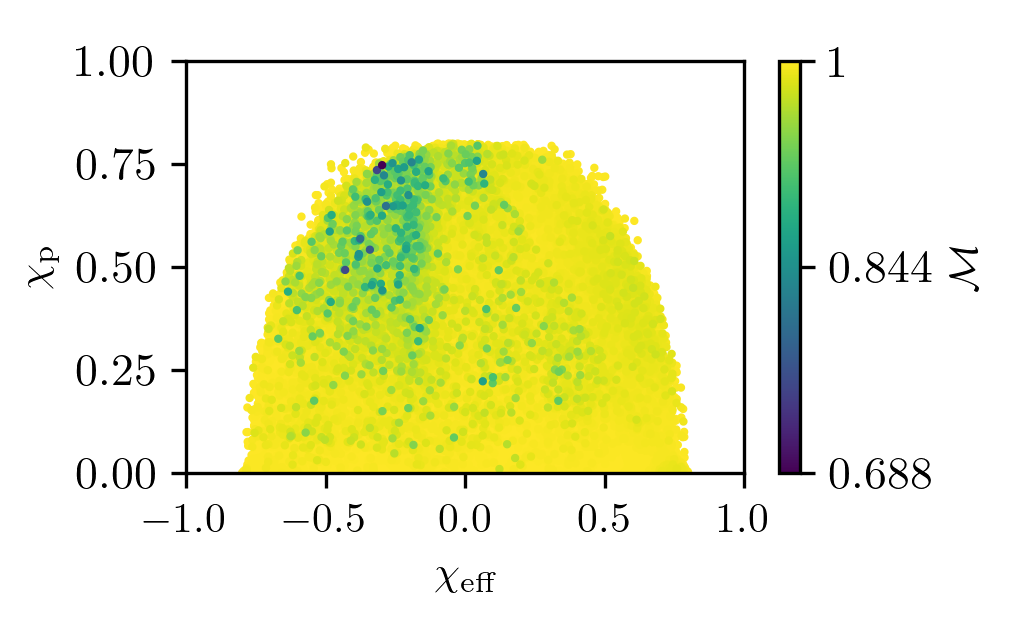}
   \caption{Scatter plots of matches between the surrogate and underlying precessing EOB waveform model
     for the validation set and a total mass of 200 $\msun$.
     The setup of the panels and the shown matches follows Fig.~\ref{fig:train_match}.
  }
  \label{fig:valid_match_200}
\end{figure*}

In this section, we discuss the mismatch results for the training and validation sets with the total mass grid of (50, 75, 100, 150, and 200) $\msun$ and inclination grid of (0, 60, 90) degrees as presented in Figs ~\ref{fig:all_mismatch_mtot_incl_median}, \ref{fig:all_mismatch_incl60} and \ref{fig:all_match_histogram}.

We show the distribution of mismatches computed using min-max match for combined training and validation configurations as a function of total mass and inclination in fig~\ref{fig:all_mismatch_mtot_incl_median}. Circles show the median of the distribution which is $\le {10}^{-3}$ which is almost an order of magnitude better for the bulk of the cases than the accuracy of 1\% we are aiming for. The 99th percentile of the distribution is shown by the downward facing triangles which show a long tail of poor matches, which goes up for high large mass. Fig.~\ref{fig:all_mismatch_incl60} shows a similar mismatch distribution for a fixed inclination where we further quantify the effects of the choice of the conservative min-max matches against optimistic max-max matches. To present the results, we decided to use the conservative min-match match method. The histograms in Fig~\ref{fig:all_match_histogram} show mismatches for various total masses at the fixed inclination of $60^{0}$.

It is important to compare the mismatch of the surrogate against the mismatches obtained for the underlying \seob~\cite{Ossokine:2020kjp} model against \ac{NR} simulations. At an inclination of 60 degrees and a total mass of $200 M_\odot$ Fig.~\ref{fig:all_mismatch_mtot_incl_median} shows a median mismatch less than $0.07\%$ and a 99th percentile mismatch of $3.5\%$. In contrast, SEOBNRv4PHM exhibits a median mismatch of $0.7\%$ against NR and a mismatch of $3\%$ at the 95th percentile (see Fig. 12 of~\cite{Ossokine:2020kjp}). Thus, we find that the surrogate has median mismatches better by one order of magnitude than the mismatches of the original model SEOBNRv4PHM against NR. In the tails of the mismatch distributions the surrogate reaches a value of $3.5\%$ at 99th percentile, only slightly above the mismatch at the 95th percentile of SEOBNRv4PHM vs NR, which again indicates the accuracy of the surrogate. The mismatches computed for the surrogate include tens of thousands of configurations and are therefore more thorough in exploring the parameter space, as well as including configurations up to mass-ratio 1:20 compared to mass-ratio 1:6 for the $\sim 1500$ precessing NR waveforms used to validate SEOBNRv4PHM. In summary, the surrogate is clearly more accurate than the underlying SEOBNRv4PHM model.

We show the results for the smallest and largest of the total masses ($50 \msun$ and $200 \msun$) for the fixed inclination of $60^{0}$ for the training set in Figs.~\ref{fig:train_match}, \ref{fig:valid_match},
\ref{fig:train_match_200}, and~\ref{fig:valid_match_200} respectively. In each of these figures, we show
scatter plots in mass-ratio, effective aligned and precession spin parameters to better understand
the distribution of matches over the parameter space and thus learn about the accuracy of the
surrogate over parameter space.

We start with Fig.~\ref{fig:train_match} which shows the matches between the surrogate and the
training set of \seob waveforms at a total masses of $50 \msun$. We see that the surrogate is in
good agreement with the parent model, with the bulk of the matches being better than 0.99 (or,
equivalently, a mismatch of less than $1\%$). We find that only  68 configurations ($\lesssim 0.1\%$) in the training set exceed this threshold.
We observe that configurations with high mismatches tend to have large effective precessional spin
parameter $\chip$, slightly anti-aligned $\chie$, and cluster towards more unequal mass-ratios $q$.
This indicates that the strongly precessing configurations are the most difficult to accurately
model, and we discuss possible explanations for this in the conclusion.
Matches calculated on the independent validation set shown in Fig.~\ref{fig:valid_match} closely
parallel the result we observed on the training set. We find 53 configurations which exceed
1\% mismatch and overall the largest values lie around 2\% as shown in Fig.~ \ref{fig:all_match_histogram}.

At higher total mass the match computation focuses on higher frequency content in the surrogate.
We have already remarked above, that surrogate fits tend to have lower accuracy for time nodes
that are close to the merger and thus surrogate accuracy worsens towards higher frequencies.
Therefore, we expect that matches will overall worsen for higher mass systems. This can be seen in
Fig.~\ref{fig:train_match_200} which corresponds to a total mass of $200 \msun$. For
this total mass, very few \ac{GW} cycles are in the \LIGO band and they are concentrated in the
merger-ring-down region. We see that the distribution pattern of the matches is similar to the $50
\msun$ result, but there is a long tail of low matches: we find that 4 \% of the cases
exceed 1\% mismatch and 70 cases have more than 10\% mismatch. Again, the behavior on the
validation set shown in Fig.~\ref{fig:valid_match_200} follows the training set very closely,
finding 4\% cases with a mismatch exceeding 1\% and 64 cases with a mismatch greater than 10\% as can be seen in Fig.~ \ref{fig:all_match_histogram}.

We further test the validity of the surrogate model for extrapolated spin magnitudes exceeding the
limit of $0.8$ we used in the surrogate construction. To do this, we compare the surrogate to the
\seob waveforms with at least one of the component spin magnitudes in the range $(0.8, 0.99)$ for a
total mass of $50 \msun$ with inclination of $60^{0}$. For spin magnitudes up to $0.9$, we obtain mismatches up to 5\%. If
we further increase the spin on the larger \ac{BH} beyond $0.9$ the mismatches can reach values as
high as 20\% if the mass ratio is larger than 5. For mass ratios less than 5, mismatches do
not exceed 10\%. In addition, mismatches do not exceed 10\% for any value of the secondary spin
magnitude if the spin magnitude of the primary is below $0.9$ -- indeed we expect that the spin of
the secondary \ac{BH} should only have a very small effect on the waveform if the mass ratio is
very unequal.


\subsection{Speed} 
\label{sub:speed}

Our main purpose of building a surrogate for \seob is to improve the speed of waveform generation
while preserving accuracy, so that it can be used for characterizing BBH mergers. In this section
we focus on the achieved speedup between the \seob model and our surrogate and compare it to one of
the efficient phenomenological waveform models which describes waveforms with similar physical
assumptions.

We compare the surrogate waveform generation time against that of \seob and the
\texttt{IMRPhenomXPHM}~\cite{Pratten:2020ceb} waveform model for different mass ratios given a fiducial
spin configuration. We choose a starting frequency of $25$ Hz. As our surrogate is only $5000$ M
long, it may not be able to start from 25 Hz for a few low-mass and asymmetric mass
configurations. In that case, we use the lowest allowed surrogate frequency to generate waveforms
and compare the costs. The total cost we consider includes the computational cost tapering and
Fourier transform (FFT) for the surrogate and \seob as both of these models are time domain.

\begin{figure}[htbp]
  \includegraphics[width=.49\textwidth]{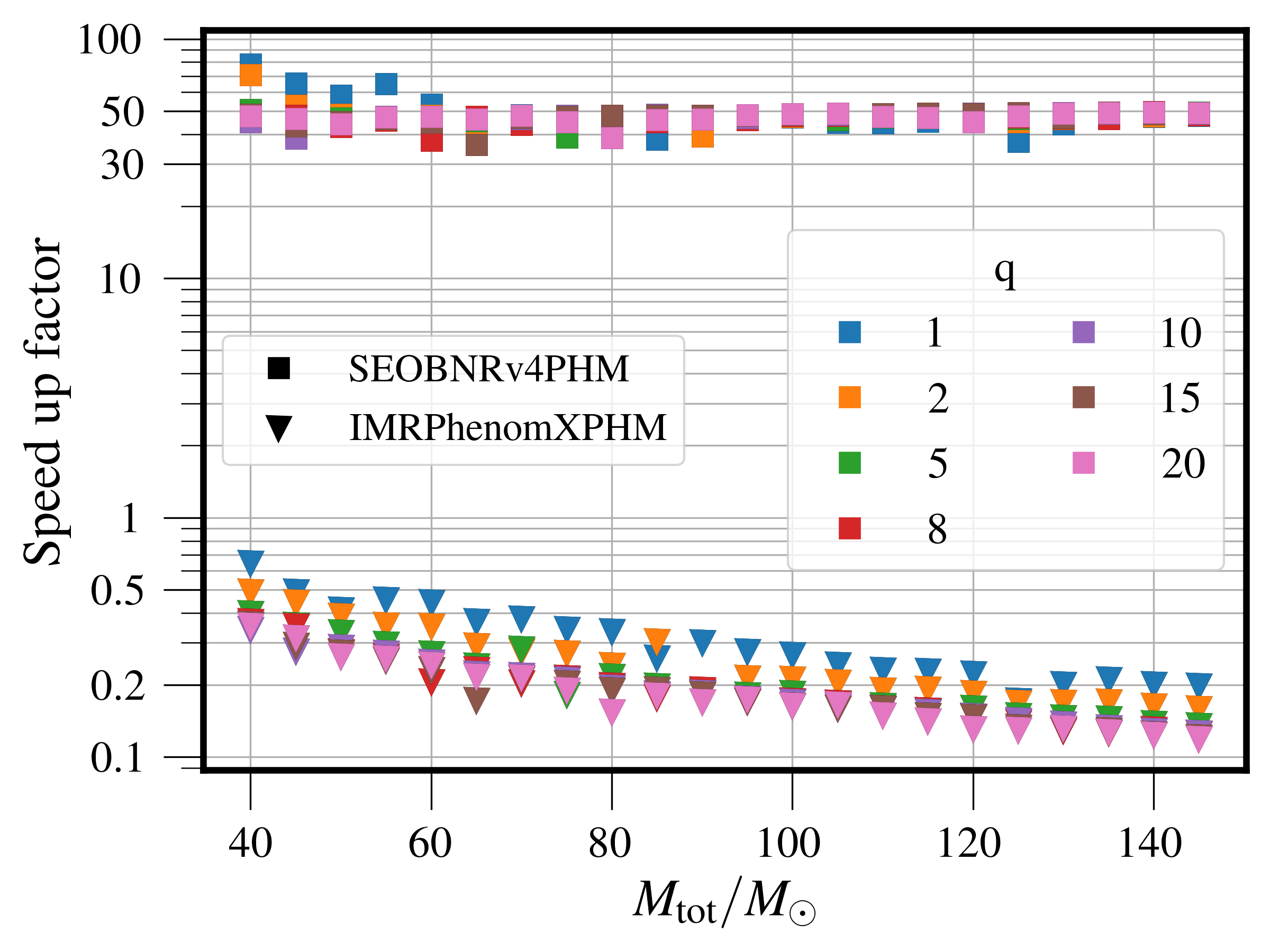}
  \caption{This figure shows the speed-up factor between
    the precessing \seob model and its surrogate (squares),
    and between the \texttt{IMRPhenomXPHM} waveform model and the \seob surrogate (triangles).
    The fiducial starting frequency is 25 Hz (see text for the details) with fixed spin vectors
    $\vec{\chi_1} = (0.6, 0.1, 0.2)$ and $\vec{\chi_2} = (0.1, 0.1, 0.6)$.
  }
   \label{fig:timing_plots}
\end{figure}

In Fig.~\ref{fig:timing_plots} we plot the ratio of the waveform generation times between either
\seob or \texttt{IMRPhenomXPHM} and the surrogate as a function of total mass and mass ratio. We can see
that the surrogate is about 40-70 times faster than the original \seob model and just 5-8 times
slower than the frequency domain phenomenological \texttt{IMRPhenomXPHM} waveform. This makes it feasible
to do Bayesian parameter estimation studies which we describe in the next section.

\subsection{Application: parameter estimation} 
\label{sub:application_parameter_estimation}

As a practical application of the new surrogate, we perform Bayesian inference on GW signals.
We consider two cases, (a) recovering a software injection of an \seob signal assuming
an average (zero) noise realization, and (b) performing inference on real data for GW150914.

For case (a) we prepared the software injection using pycbc~\cite{pycbc_2021_5347736}.
The signal has a detector-frame total mass $70 M_\odot$, (inverse) mass-ratio $\bar q := 1/q = 1/5$,
dimensionless component spins of $\vec\chi_1 = (0.5, 0.05, 0.3)$ and $\vec\chi_2 = (0.3, -0.2, 0.3)$ at
a reference frequency of $20$ Hz. The inclination angle of the signal is
$\theta_\mathrm{JN} = 2.58$.
The chosen signal lies in the vicinity of subdomain boundaries ($q=4$, $\chi_\mathrm{eff} = 0.3$)
of the surrogate model. Therefore, it constitutes a test of the smoothness of the surrogate at
these boundaries where we switch between independently trained surrogates defined on adjacent
subdomains (see Fig.~\ref{fig:subdomains}). The signal has a network SNR of $\sim 27$ in
the three detector advanced LIGO - Virgo HLV network.
We use the bilby Bayesian inference code~\cite{bilby_2019ApJS..241...27A} with the
dynesty~\cite{2020MNRAS.493.3132S} nested sampler~\cite{2004AIPC..735..395S, 10.1214/06-BA127}.

\begin{figure*}[htbp]
  \centering
  \includegraphics[width=.45\textwidth]{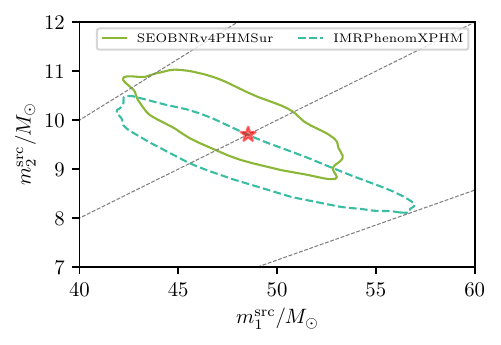}
  \includegraphics[width=.45\textwidth]{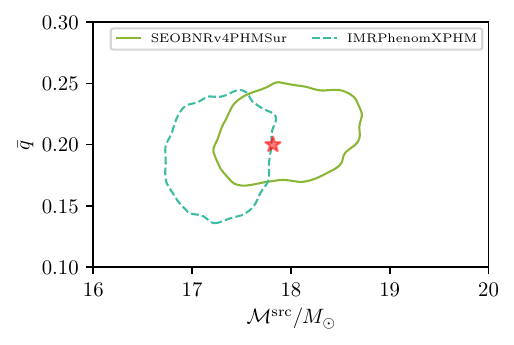}
  \includegraphics[width=.45\textwidth]{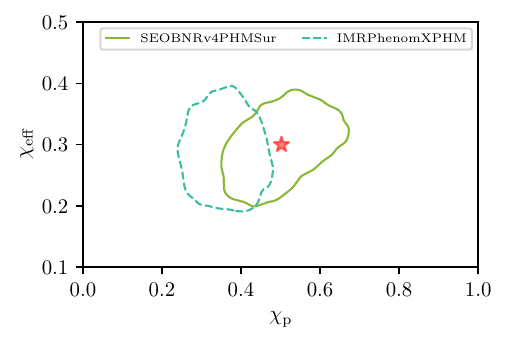}
  \includegraphics[width=.45\textwidth]{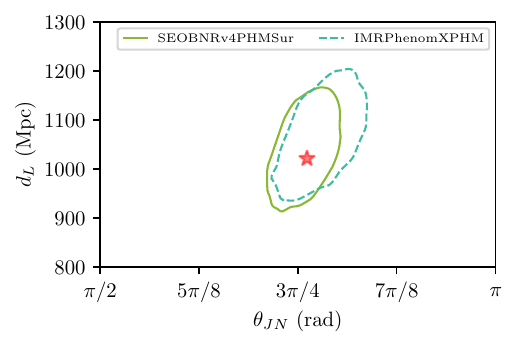}
  \caption{
  Marginal posterior PDFs for \texttt{SEOBNRv4PHMSur} and \texttt{IMRPhenomXPHM} on a zero-noise \seob signal.
  The panels show 90\% credible regions for the following quantities
  \emph{top left}: source-frame component masses (and lines of constant (inverse) mass-ratios $\bar q$: 1/4, 1/5, 1/7),
  \emph{top right}: the source-frame chirp mass and (inverse) mass ratio,
  \emph{bottom left}: the effective aligned and effective precessing spin parameters, and
  \emph{bottom right}: the luminosity distance and inclination angle.
  }
  \label{fig:PE_plots_q5inj}
\end{figure*}

Marginal posterior PDFs are shown in Fig.~\ref{fig:PE_plots_q5inj} where the signal is
recovered with the surrogate and, as a comparison, with \texttt{IMRPhenomXPHM}~\cite{Pratten:2020ceb}.
Since we are using a zero-noise injection we expect that the likelihood peaks close to
the signal parameters, and since the signal is loud, this should also be the case for the
posterior distribution. We can verify that indeed the contours indicating the 90\% credible
regions of the posterior PDFs are close to centered around the true signal parameters indicated
by red asterisks for all parameters shown. We do not see any indication that the posterior
would be non-smooth at subdomain boundaries which is expected given the demonstrated accuracy
of the surrogate compared to the \seob model.
The \texttt{IMRPhenomXPHM} posteriors show a noticeable bias in chirp mass and effective precession
spin, with the true parameter values close to 90\% credible region boundary. This is not too
surprising, since \seob and \texttt{IMRPhenomXPHM} are expected to differ at the level of
up to a few percent in mismatch (see Fig. 9 in Ref.~\cite{Colleoni:2020tgc}).
Specifically, for the injection we have considered, the match (starting from 20 Hz) between
\seob and its surrogate is 0.997 while the match between \seob and \texttt{IMRPhenomXPHM} is 0.96.
Smaller discrepancies are seen in distance and inclination.
We can also see that the shape and extent of the 90\% credible regions are overall similar,
but, for instance, the mass-ratio is estimated more precisely for the surrogate compared to
\texttt{IMRPhenomXPHM}.

In case (b) we analyzed GW150914 with the surrogate model and with SEOBNRv4PHM.
For LVK analyses of this event see~\cite{LIGOScientific:2016vlm,PhysRevX.9.031040,LIGOScientific:2021usb}.
Our analysis of GW150914 used the parallel version of the bilby code~\cite{Smith:2019ucc}
and open data from GWOSC~\cite{LIGOScientific:2019lzm}.
The parameter estimation run with the surrogate model took
18 hours on 144 cores~\footnote{We have used the following dynesty sampler settings:
nlive=2000, nact=20, and a dlogz stopping criteria of 0.1.}.
The corresponding analysis with SEOBNRv4PHM took 52 hours on 1024 cores, or 15 days on 144 cores,
assuming perfect scaling, a speedup of 20.
Because the surrogate only supports spin magnitudes up to 0.8 we apply the same restriction to SEOBNRv4PHM.

\begin{figure*}[htbp]
  \centering
  \includegraphics[width=.45\textwidth]{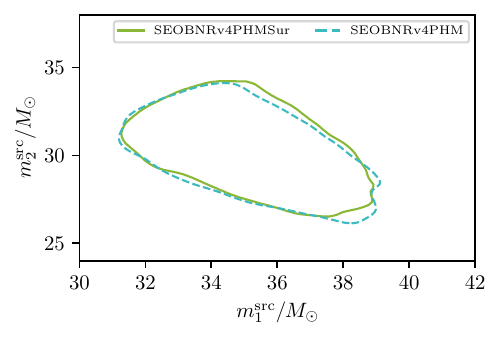}
  \includegraphics[width=.45\textwidth]{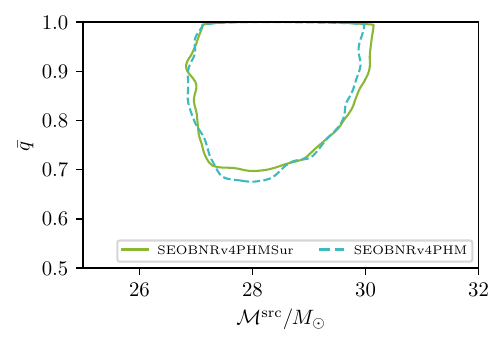}
  \includegraphics[width=.45\textwidth]{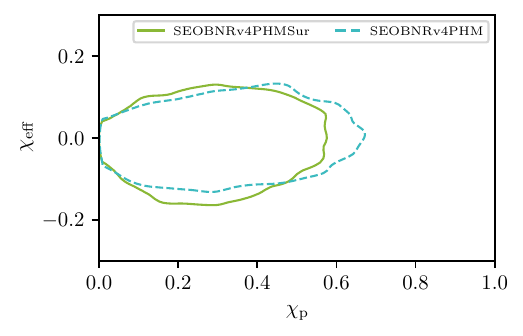}
  \includegraphics[width=.45\textwidth]{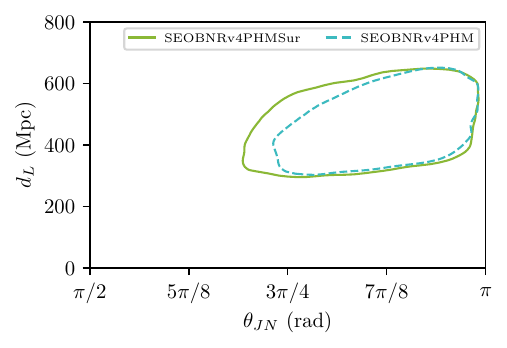}
  \caption{Marginal posterior PDFs for \texttt{SEOBNRv4PHMSur} and \seob for GW150914.
  The panels show 90\% credible regions for the following quantities
  \emph{top left}: source-frame component masses,
  \emph{top right}: the source-frame chirp mass and (inverse) mass ratio,
  \emph{bottom left}: the effective aligned and effective precessing spin parameters, and
  \emph{bottom right}: the luminosity distance and inclination angle.
  }
  \label{fig:PE_plots_GW150914}
\end{figure*}

We see in Fig.\ref{fig:PE_plots_GW150914} that agreement between the marginal posteriors for the two models
is fairly good, but there are some visible deviations. Quantitatively, the Jensen-Shannon divergence (JSD)
between the 1-dimensional marginal posteriors is smaller than 0.02 bits, with the largest values occurring
for the chirp mass, the mass of the secondary, and $\chi_\mathrm{p}$. Such differences can arise from the
lower accuracy of the surrogate model for large values of the effective precession spin. In this mass range
the mismatch can exceed $1\%$ for 4\% of the cases shown in Fig.~\ref{fig:all_match_histogram}.

\section{Conclusion}
\label{sec:conclusion}

In this paper, we have presented a surrogate for the precessing \seob waveform model
based on techniques developed for NR surrogates with a few modifications.
In particular, we have used domain decomposition (see Sec.~\ref{sub:domain_decomposition})
to deal with the large range the surrogate covers in mass ratio while keeping
the polynomial fits below cubic order.
In Sec.~\ref{sub:training_set_generation} we discussed
that a larger training set along with an enrichment strategy is beneficial to achieve good
accuracy for at least some of the subdomains. Over all domains, we used a dataset of
about 170,000 waveforms in total when combining the training and validation sets
which is about two orders of magnitude more waveform data than used for NR surrogates so far.
This surrogate is built up to mass ratio $q=20$ and spin magnitudes of $0.8$.
When we attempted to build a surrogate for higher spin magnitudes we found that the
fit accuracy degraded substantially and we leave such an extension for future work.

As has been observed previously for NR surrogates~\cite{Varma:2019csw} we find that the accuracy
of our surrogate is limited by the modeling of the orbital phase (or frequency) as shown in
Figs.~\ref{fig:surrogate_train_errors} and~\ref{fig:surrogate_test_errors}. We observe that the
accuracy of the fits decreases as we approach the merger, presumably because the waveforms are
less smooth in this regime.
As shown in Figs.~\ref{fig:train_match} and~\ref{fig:valid_match} we find that the bulk of the mismatches between the surrogate and \seob
is smaller than 1\% with the largest values around 2\% for a total mass of $50 \msun$. Since the surrogate fits worsen as we approach the merger, the surrogate accuracy degrades for systems of high total masses.
At a total mass of $200 \msun$ we find that 4\% of the configurations exceed a mismatch of 1\%.
The worst matches are found at large values of the effective precession spin parameter $\chi_\mathrm{p}$, slightly negative effective aligned spin $\chi_\mathrm{eff}$, and mass ratios greater than 4.
The surrogate's reduced performance in this region of parameter space could be due
to any combination of the modeling approximations or the underlying \seob waveform model itself. To
check the former, we carried out extensive experimentation with parameters
defining the surrogate model (fit tolerances, the data pieces' parameterizations, number of basis elements, number of dynamics nodes, etc). To check the \seob waveform data we experimented with turning off the non-quasi-circular orbit corrections and checking non-smoothness diagnostics. Despite these efforts, we have been unable to determine the underlying cause of this problem.

While it is possible to evaluate the surrogate outside of its training domain, such as
extrapolating the surrogate to higher spin magnitudes than the training bound of $0.8$, this
incurs an increase of mismatches up to 5\% ($a < 0.9$) and 20\% ($a < 0.99$ and $q > 5$) for q
total mass of $50 \msun$.

In Fig.~\ref{fig:timing_plots} we demonstrate that our surrogate is about 50 times faster
than \seob and about 5 times slower than the \texttt{IMRPhenomXPHM} waveform model.
This is a significant speedup compared to the original \ac{EOB} model and will enable
the use of standard sampling methods for Bayesian inference with this waveform family.

As an application of our surrogate to data analysis studies we showed in
Fig.~\ref{fig:PE_plots_q5inj} that it allows for unbiased estimation of binary parameters
when the precessing GW signal is modeled by \seob and we are using an average ``zero''
noise realization. The surrogate also leads to posterior distributions which are consistent
with, albeit slightly broader, than those found for SEOBNRv4PHM of GW150914 as
demonstrated in Fig.~\ref{fig:PE_plots_GW150914}.

The surrogate we present here has a duration of 5000 M and cannot model
the inspiral for low-mass binaries. Even though a computationally efficient
post-adiabatic processing model from the TEOB family of arbitrary duration is
available (see Ref~\cite{Gamba:2021ydi} and ~\cite{Gamba:2020ljo}), it is still of interest to build
surrogates for significantly longer waveforms to enable an accelerated analysis
of costly low-mass GW events and to understand waveform systematics. The first
step is to investigate the limitations of the current surrogate construction
method.
To this end, we also constructed a longer surrogate (covering 80000 M), which corresponds
to a starting frequency of 20 Hz for an equal mass binary with a total mass of $20 \msun$.
This surrogate only includes two subdomains in mass ratio with the boundaries at $q = 1, 2,$
and 4. We found a surrogate accuracy similar to the shorter surrogate we have discussed in-depth
in this paper when restricting to the overlapping region of parameter space.
This longer surrogate turns out to be more than an order of magnitude faster computationally
than \seob.
We found that the construction of this longer surrogate for each of its subdomains is
computationally prohibitive when using 25 points per orbit to construct the dynamical
surrogate which would result in 20000 time points.
Apart from this computational issue, we did not find any hard limitations in the current method,
in particular its reliance on the solution of an ODE system in time and the fitting errors.
Yet, even a duration of 80000 M is not long enough in practice as the duration increases
very rapidly as the total mass and/or the starting frequency decreases, and for low-mass
binaries detectable by ground-based detectors, we would need to cover durations of millions of M,
which does not seem feasible with the current methodology.
Therefore, it will be very interesting to study alternative surrogate construction methods or
hybridization of the surrogate with post-Newtonian or post-adiabatic \ac{EOB} waveforms in the
inspiral.

\section*{Acknowledgements}

We thank Jonathan Blackman, Alessandra Buonanno, Steffen Grunewald, Sylvain Marsat, Ajit Mehta, and Harald Pfeiffer for helpful discussions throughout the project. We like to thank Marta Colleoni for carefully going through the manuscript and providing useful comments.
This research has made use of data, software and/or web tools obtained from the Gravitational Wave Open Science Center (https://www.gw-openscience.org/ ), a service of LIGO Laboratory, the LIGO Scientific Collaboration and the Virgo Collaboration.
S.E.F. acknowledges support from NSF Grants Nos.~PHY-2110496 and PHY-1806665.
This project has received funding from the European Union’s Horizon 2020
research and innovation program under the Marie Skłodowska-Curie grant
agreement No.~896869.
V.V.~acknowledges support from NSF Grant No. PHY-2309301 and UMass Dartmouth's
Marine and Undersea Technology (MUST) Research Program funded by the Office of
Naval Research (ONR) under Grant No. N00014-23-1–2141.
A significant part of the computational work for the project is done on Hypatia, a general-purpose gravitational wave data analysis cluster and a couple of large workstations named Saga and Odin at AEI Potsdam-Golm.
A portion of the computational work of this project was performed on the CARNiE cluster
at UMassD, which is supported by the ONR/DURIP Grant No.\ N00014181255.
LIGO Laboratory and Advanced LIGO are funded by the United States National Science Foundation (NSF) as well as the Science and Technology Facilities Council (STFC) of the United Kingdom, the Max-Planck-Society (MPS), and the State of Niedersachsen/Germany for support of the construction of Advanced LIGO and construction and operation of the GEO600 detector. Additional support for Advanced LIGO was provided by the Australian Research Council. Virgo is funded, through the European Gravitational Observatory (EGO), by the French Centre National de Recherche Scientifique (CNRS), the Italian Istituto Nazionale di Fisica Nucleare (INFN) and the Dutch Nikhef, with contributions by institutions from Belgium, Germany, Greece, Hungary, Ireland, Japan, Monaco, Poland, Portugal, Spain. This manuscript has been assigned a LIGO Document No LIGO-P2200040.

\bibliography{references}

\end{document}